\documentclass[12pt,letterpaper]{article}

\usepackage[utf8]{inputenc}
\usepackage[T1]{fontenc}
\usepackage{lmodern}

\usepackage{microtype}
\microtypesetup{protrusion=true, expansion=true, nopatch=footnote}

\usepackage{geometry}
\usepackage{amsmath,amssymb,amsthm}
\usepackage{siunitx}

\usepackage{graphicx}
\usepackage[table]{xcolor}
\usepackage{array,booktabs,ragged2e,tabularx,longtable,ltablex,makecell}
\keepXColumns
\usepackage{adjustbox}

\usepackage{tikz}
\usepackage{pgfplots}
\pgfplotsset{compat=1.18}
\usetikzlibrary{arrows.meta, positioning, calc, decorations.pathreplacing}

\usepackage{caption}
\usepackage{subcaption}
\usepackage[most]{tcolorbox}
\usepackage{enumitem}
\usepackage{placeins}
\usepackage{float}

\usepackage{textcomp}
\usepackage{xurl}
\usepackage[numbers,sort&compress]{natbib}
\usepackage{doi}

\allowdisplaybreaks
\DeclareUnicodeCharacter{2212}{-}
\DeclareUnicodeCharacter{2013}{--}
\DeclareUnicodeCharacter{2014}{---}
\DeclareUnicodeCharacter{2261}{\ensuremath{\equiv}}
\DeclareUnicodeCharacter{2248}{\ensuremath{\approx}}
\DeclareUnicodeCharacter{03B1}{\ensuremath{\alpha}}
\DeclareUnicodeCharacter{03B2}{\ensuremath{\beta}}

\newcolumntype{L}[1]{>{\RaggedRight\arraybackslash}p{#1}}
\newcolumntype{C}[1]{>{\centering\arraybackslash}p{#1}}
\newcolumntype{Y}{>{\RaggedRight\arraybackslash}X}

\usepackage{titlesec}
\titleformat{\paragraph}[block]{\normalfont\normalsize\bfseries}{}{}{ }
\titlespacing*{\paragraph}{0pt}{*1.5}{0.5ex}
\usepackage{hyperref}
\usepackage{bookmark}

\graphicspath{{figs/}{figures/}}

\hypersetup{
	colorlinks=true,
	linkcolor=blue,
	citecolor=blue,
	urlcolor=blue,
	pdftitle={Empirical Universal Scaling of Neutron-Skin Curvature Across the Nuclear Chart},
	pdfauthor={Brent Baker},
	pdfsubject={Empirical neutron-skin scaling; curvature normalization},
	pdfkeywords={neutron skin, nuclear radii, curvature ratio, Compton scale, isotopes, scaling law},
	pdfproducer={LaTeX},
	pdfcreator={Brent Baker (Paper 1: Empirical)}
}

\title{%
	Empirical Universal Scaling of Neutron-Skin Curvature\\
	Across the Nuclear Chart
}
\author{%
	Brent Baker\thanks{%
		Correspondence: \texttt{brentbaker0918@gmail.com}.\\
		ORCID: \texttt{0009-0000-2115-3573}
	}
}
\date{\today}

\begin{document}
	\maketitle
	
	\begin{abstract}

	Neutron-rich surface structure---often quantified by the neutron-skin thickness
	$\Delta r_{np}=r_n-r_p$---encodes essential information about nuclear geometry,
	surface structure, and isovector response, yet a compact, universal description
	across the nuclear chart has remained elusive. In this work we present an 
	empirical analysis of charge-radius–derived neutron-excess surface response, 
	using a mass-normalized curvature framework constructed directly from 
	experimental charge radii. Here	``neutron-skin curvature'' denotes an 
	\emph{operational, charge-radius--derived proxy} for neutron-excess surface 
	growth and should not be conflated with $\Delta r_{np}$ extracted from 
	neutron-sensitive probes.
	
	By normalizing nuclear radii to a relativistic quantum length scale proportional
	to $\hbar/(mc)$, we construct a dimensionless curvature ratio that enables
	direct comparison across isotopic chains of widely varying mass. When the
	resulting skin-curvature proxy is expressed as a function of normalized neutron
	excess, data from more than 800 nuclei spanning 88 elements collapse onto a
	single universal curve without element-specific rescaling or interaction-model
	tuning (the reference curve is used only as a fixed empirical baseline for 
	residual analysis). This collapse accounts for approximately $88\%$ of the
	observed variance and achieves a substantially tighter cross-element collapse
	than standard droplet-style baselines fitted to the same dataset.
	
	Analysis of the residuals reveals structured, physically meaningful deviations
	rather than random scatter. Three finite-size regimes are identified: an initial
	skin-formation regime at low neutron excess, a relaxation regime in which
	nuclei converge toward a bulk curvature geometry, and a saturation regime at
	higher neutron excess. Very light nuclei ($Z \le 4$) occupy a distinct few-body
	domain and are shown to lie outside the bulk scaling regime. Stratification by
	conventional periodic-table families further reveals tighter sub-manifolds for
	several groups, indicating additional geometric constraints superimposed on the
	universal trend.
	
	All results reported here arise directly from experimental data and standard
	physical constants, without introducing new interaction terms or modifying
	established nuclear theory. The observed universal scaling establishes a robust
	empirical foundation for charge-radius--derived isovector surface geometry and
	motivates further investigation of its consequences in atomic and molecular
	systems.
\end{abstract}
	
	\clearpage
	\setcounter{tocdepth}{2}
	\setcounter{secnumdepth}{3}
	\tableofcontents
	\clearpage
	
	\section{Introduction}
\label{sec:introduction}
Neutron-rich surface structure---often quantified by the neutron-skin thickness
$\Delta r_{np}=r_n-r_p$---provides a sensitive probe of nuclear geometry, surface
structure, and isovector behavior. Differences between neutron and proton
distributions influence a wide range of observables, including charge radii,
separation energies, collective excitations, and parity-violating electron
scattering results\cite{PREXII2021,CREX2022}. In this paper we do not attempt to extract $r_n$ (and thus
$\Delta r_{np}$) directly; instead, we introduce a charge-radius--derived,
dimensionless proxy (``neutron-skin curvature'') designed to organize
neutron-excess surface systematics across the nuclear chart.

Despite extensive experimental and theoretical work, neutron-skin behavior is most commonly analyzed within individual isotopic chains or model-specific frameworks. Charge radii are typically examined as functions of neutron number, mass number, or $A^{1/3}$ scaling, while theoretical interpretations often rely on symmetry-energy parameters within energy-density-functional or droplet-model descriptions\cite{Krane1987,Horowitz2014,Bender2003,MyersSwiatecki1969}. These approaches are effective for reproducing local trends but do not naturally yield a compact, cross-element organization of neutron-skin geometry across the nuclear chart.

A central difficulty is that nuclear size varies strongly with mass. When radii are compared in physical units, geometric structure is entangled with trivial mass dependence, obscuring potential dimensionless regularities. As a result, even high-quality experimental datasets appear fragmented when viewed globally, and systematic comparison between elements becomes challenging.

In this work we adopt a deliberately empirical approach aimed at isolating geometric structure from mass scaling. By expressing experimental charge radii in terms of a mass-based, dimensionless curvature ratio and decomposing that ratio into core and neutron-excess surface (``skin'') contributions, we uncover a universal organization of neutron-skin geometry that is not apparent in conventional representations. This organization emerges directly from evaluated nuclear data and does not require model-dependent assumptions, fitted interaction terms, or element-specific adjustments.

The analysis presented here addresses a different question than those typically posed by detailed nuclear models. Rather than seeking to optimize the description of individual nuclei, we ask whether neutron-skin behavior exhibits a common geometric structure when expressed in appropriate normalized variables. As shown below, this approach reveals a strong collapse of neutron-skin curvature across hundreds of nuclei and exposes structured residual behavior associated with finite-size effects, shell structure, and family-level trends. Throughout this work, the term “curvature” is used strictly as a geometric descriptor derived from experimental radii and mass-based normalization. Likewise, ``neutron-skin curvature'' refers to the above charge-radius--derived proxy and should not be conflated with the neutron-skin thickness $\Delta r_{np}$ extracted from neutron-sensitive probes. No assumptions are made regarding underlying nuclear forces or microscopic dynamics. Interpretive frameworks that may motivate or contextualize this geometric organization are intentionally deferred to a companion theoretical study.

The paper is organized as follows. In Sec.~\ref{sec:Mass-Based Radius Normalization}, we introduce a mass-based normalization that places nuclear radii on a common scale. Sec.~\ref{sec:Core--Skin Decomposition} defines a geometric core--skin decomposition used throughout the analysis. The resulting universal neutron-skin scaling is presented in Sec.~\ref{sec:universal_skin_scaling} and compared to standard baselines in Sec.~\ref{sec:baseline_comparison}. Residual structure and finite-size regimes are examined in Sec.~\ref{sec:residual_regimes}, followed by a discussion of domain-of-validity limits for light nuclei in Sec.~\ref{sec:light_nucleus_validity}. Sec.~\ref{sec:family_conditioned_structure} explores family-conditioned residual behavior. We conclude with a synthesis of results and their implications in Secs.~\ref{sec:discussion} and~\ref{sec:conclusion}.

	\section{Mass--Based Radius Normalization}
\label{sec:Mass-Based Radius Normalization}
Neutron-rich surface structure---often quantified by the neutron-skin thickness
$\Delta r_{np}=r_n-r_p$---provides a sensitive probe of nuclear geometry, surface
structure, and isovector behavior. Differences between neutron and proton
distributions influence a wide range of observables, including charge radii,
separation energies, collective excitations, and parity-violating electron
scattering results. In this paper we do not attempt to extract $r_n$ (and thus
$\Delta r_{np}$) directly; instead, we introduce a charge-radius--derived,
dimensionless proxy (``neutron-skin curvature'') designed to organize
neutron-excess surface systematics across the nuclear chart.

To separate geometry from mass, we introduce a dimensionless normalization based on the standard energy--frequency relation
\begin{equation}
	E = h\nu,
\end{equation}
together with the relativistic mass--energy equivalence
\begin{equation}
	E = mc^2.
\end{equation}
Combining these identities yields a mass-associated frequency scale
\begin{equation}
	\nu = \frac{mc^2}{h},
\end{equation}
and the corresponding reduced Compton wavelength, the natural mass-linked length scale formed from $\hbar$, $m$, and $c$,
\begin{equation}
	\bar{\lambda}_C \equiv \frac{\hbar}{mc}.
\end{equation}.

The quantity $\bar{\lambda}_C$ is the reduced Compton wavelength associated with mass $m$.
In the present work, it is employed solely as a normalization scale derived from standard physical constants. 
No new physical assumption is introduced by this construction; it serves only to provide a common mass-dependent reference length.

While the choice of $\bar{\lambda}_C$ is not arbitrary, a detailed physical interpretation of this scale is not required to establish the empirical scaling reported here. 
The results presented in this paper depend only on the role of $\bar{\lambda}_C$ as a consistent mass-linked normalization, not on any particular microscopic or dynamical explanation. 
A more complete discussion of the physical motivation and broader implications of this normalization is deferred to future work.

Using this scale, we define a dimensionless curvature ratio
\begin{equation}
	K_R \equiv \frac{r_{\mathrm{exp}}}{\bar{\lambda}_C},
\end{equation}
where $r_{\mathrm{exp}}$ is the experimentally measured nuclear charge radius. This ratio places radii on a common dimensionless scale, reducing trivial mass-linked scaling so that residual geometric structure can be compared across elements. All subsequent analysis is performed in terms of $K_R$ and its derived components.

We emphasize that the empirical results presented below do not depend on any particular interpretation of $\bar{\lambda}_C$ beyond its role as a mass-based normalization. As shown in the following sections, expressing nuclear radii in this dimensionless form reveals a universal scaling of the charge-radius--derived skin-curvature proxy that is not apparent when radii are analyzed in physical units alone.
	\section{Core--Skin Decomposition}
\label{sec:Core--Skin Decomposition}
Experimental charge radii measure the rms spatial extent of the nuclear charge
distribution (dominated by protons). Although neutrons carry no net charge,
changes in neutron number influence charge radii indirectly through nuclear
structure and correlations. To isolate the geometric imprint of neutron excess
in a dimensionless way, we decompose the curvature ratio $K_R$ into a reference
(core) component and an operational surface (``skin'') component using only
experimentally evaluated data.

Charge-radius values $r_{\mathrm{exp}}$ are taken from the evaluated compilation of Angeli and Marinova~\cite{AngeliMarinova2013}, while nuclear masses used in the normalization scale are taken from the Atomic Mass Evaluation (AME2020)~\cite{AME2020}. Fundamental constants are adopted from the CODATA recommendations~\cite{CODATA2018}.

For each element $Z$, we identify a reference isotope corresponding to the smallest neutron number for which reliable charge-radius data exist and for which the nucleus is long-lived. This isotope defines the \emph{core configuration} for that element. The corresponding curvature ratio is denoted
\begin{equation}
	K_{R,\mathrm{core}}(Z) \equiv K_R(Z, N_{\mathrm{core}}),
\end{equation}
where $N_{\mathrm{core}}$ is the neutron number of the reference isotope as determined directly from the evaluated datasets.  Sensitivity to this operational core-isotope choice is examined in Appendix~\ref{app:robustness_checks}.

For isotopes with additional neutrons, the total curvature ratio is written as
\begin{equation}
	K_R^2(Z,N) = K_{R,\mathrm{core}}^2(Z) + K_{R,\mathrm{skin}}^2(Z,N),
\end{equation}
which defines a charge-radius--derived skin-curvature proxy
\begin{equation}
	K_{R,\mathrm{skin}}(Z,N) \equiv \sqrt{K_R^2(Z,N) - K_{R,\mathrm{core}}^2(Z)}.
\end{equation}

This decomposition is purely algebraic: it defines $K_{R,\mathrm{skin}}$ as the
nonnegative increment in $K_R$ relative to the chosen reference isotope via an
rms (quadrature) relation. The quantity $K_{R,\mathrm{skin}}$ is therefore an
\emph{operational proxy} derived from charge-radius systematics; it should not
be conflated with the neutron-skin thickness $\Delta r_{np}=r_n-r_p$ obtained
from neutron-sensitive probes. No neutron density profile is assumed beyond
its indirect influence on the measured charge radius. The square-root form
ensures that $K_{R,\mathrm{skin}}$ vanishes identically for the core isotope and
provides a convenient magnitude for neutron-excess--driven surface response;
small nonmonotonicity can occur due to experimental uncertainties or known
isotope-shift anomalies.

We emphasize that the terms ``core'' and ``skin'' are used here as operational descriptors rather than as rigid physical partitions. The decomposition provides a convenient way to isolate the incremental geometric effect of neutron excess while maintaining a dimensionless, cross-element normalization that is robust against trivial mass scaling.

Throughout this work, the neutron excess is defined as
\begin{equation}
	N_{\mathrm{excess}} \equiv N - N_{\mathrm{core}},
\end{equation}
and isotopic trends are examined as functions of the normalized variable $N_{\mathrm{excess}}/Z$. As shown in subsequent sections, this decomposition exposes a universal scaling behavior of the skin-curvature proxy that is not apparent when radii are analyzed in physical units alone.
	
	\section{Universal Neutron--Skin Scaling}
\label{sec:universal_skin_scaling}

Using the mass-normalized curvature framework and core--skin decomposition defined above, we examine the behavior of neutron-skin curvature across the nuclear chart. For each isotope, we consider the dimensionless ratio
\begin{equation}
	y \equiv \frac{K_{R,\mathrm{skin}}}{K_{R,\mathrm{core}}},
\end{equation}
plotted as a function of the normalized neutron excess
\begin{equation}
	x \equiv \frac{N_{\mathrm{excess}}}{Z}.
\end{equation}

When $y$ is plotted against $x$ for all validated isotopes, data from disparate elements and mass ranges collapse onto a single, smooth curve. This collapse spans more than 800 nuclei across 88 elements and is observed over the domain
\[
0 \le \frac{N_{\mathrm{excess}}}{Z} \le 0.5.
\]
No element-specific rescaling, shell corrections, or fitted interaction parameters are applied.
\begin{figure}[t]
	\centering
	\includegraphics[width=0.85\linewidth]{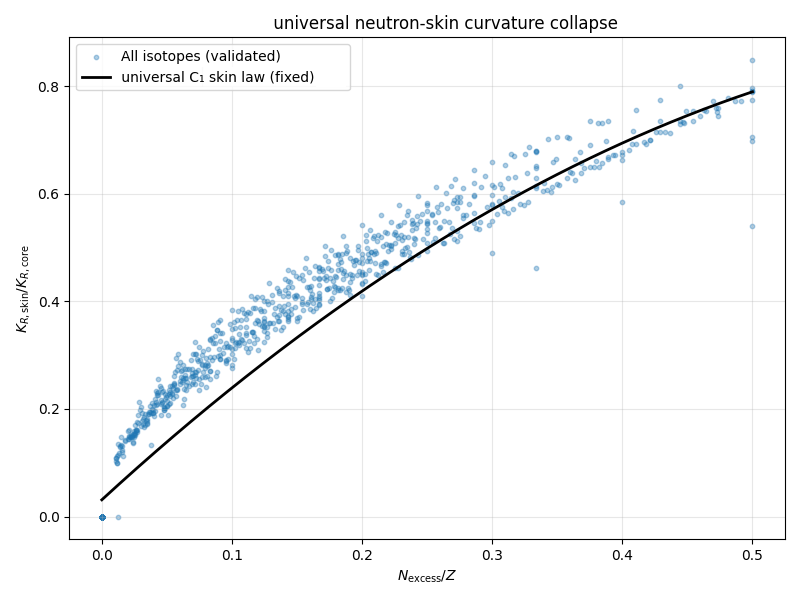}
	\caption{
		Universal neutron-skin curvature collapse across the nuclear chart.
		The dimensionless ratio $K_{R,\mathrm{skin}}/K_{R,\mathrm{core}}$ is plotted as a function of normalized neutron excess $N_{\mathrm{excess}}/Z$ for all validated nuclei.
		Data from 826 isotopes spanning 88 elements collapse onto a single curve without fitted parameters.
		The solid line shows the fixed empirical baseline $F(x)$ used for residual analysis.
	}
	\label{fig:universal_collapse}
\end{figure}

Because $K_R=r_{\mathrm{exp}}/\bar{\lambda}_C$ explicitly removes a mass-linked scale (and nuclear
masses vary systematically along isotopic chains), part of the smooth backbone collapse is
expected from the coordinate transformation itself. The nontrivial empirical result is that,
in these mass-normalized coordinates, disparate elements exhibit a tight collapse onto a single
reference curve and the remaining deviations form structured residuals rather than random
scatter (Sec.~\ref{sec:residual_regimes}). A matched-complexity comparison against radii-only
coordinates and element-wise cross-validation are given in Appendix~\ref{app:mass_scaling_test}. Expressed in this normalized form, isotopic chains that appear distinct when analyzed in physical units become mutually consistent, revealing a common geometric scaling behavior. The function $F(x)$ is used solely as a fixed reference for residual analysis and is not interpreted as a physical model.

For reference, the empirical shape of the collapsed curve is well approximated by a concave-down quadratic function of $x$ over the validated domain. The explicit functional form is reported for reproducibility but is not introduced as a fitted model in the analysis below; all comparisons and residuals are evaluated relative to this fixed baseline.

We emphasize that this scaling is empirical in nature. It arises directly from experimental charge radii and evaluated nuclear masses, combined through a mass-based normalization and geometric decomposition. No assumptions regarding nuclear forces, symmetry-energy parameters, or shell structure are required to obtain the collapse.

In the following sections, we compare this curvature-normalized scaling to standard baseline constructions and examine the structured residual behavior that remains after the universal trend is removed.
	\section{Comparison to Standard Baselines}
\label{sec:baseline_comparison}

To assess the significance of the curvature-normalized scaling presented in Sec.~\ref{sec:universal_skin_scaling}, we compare it against two baseline constructions that represent common ways neutron-skin behavior is examined using existing nuclear data. These baselines are chosen to be model-minimal and to use the same underlying experimental datasets, ensuring a fair comparison.

\subsection{Radii--Only Geometric Baseline}

A natural reference point is a radii-only construction using experimental charge radii without mass normalization. For each element $Z$, a core radius $r_{\mathrm{core}}$ is defined using the same core isotope identified in Sec.~3, and a geometric skin proxy is extracted via
\begin{equation}
	r_{\mathrm{skin}} = \sqrt{r_{\mathrm{exp}}^2 - r_{\mathrm{core}}^2}.
\end{equation}
A dimensionless quantity $r_{\mathrm{skin}}/r_{\mathrm{core}}$ is then examined as a function of $N_{\mathrm{excess}}/Z$.
\begin{figure}[t]
	\centering
	\includegraphics[width=0.85\linewidth]{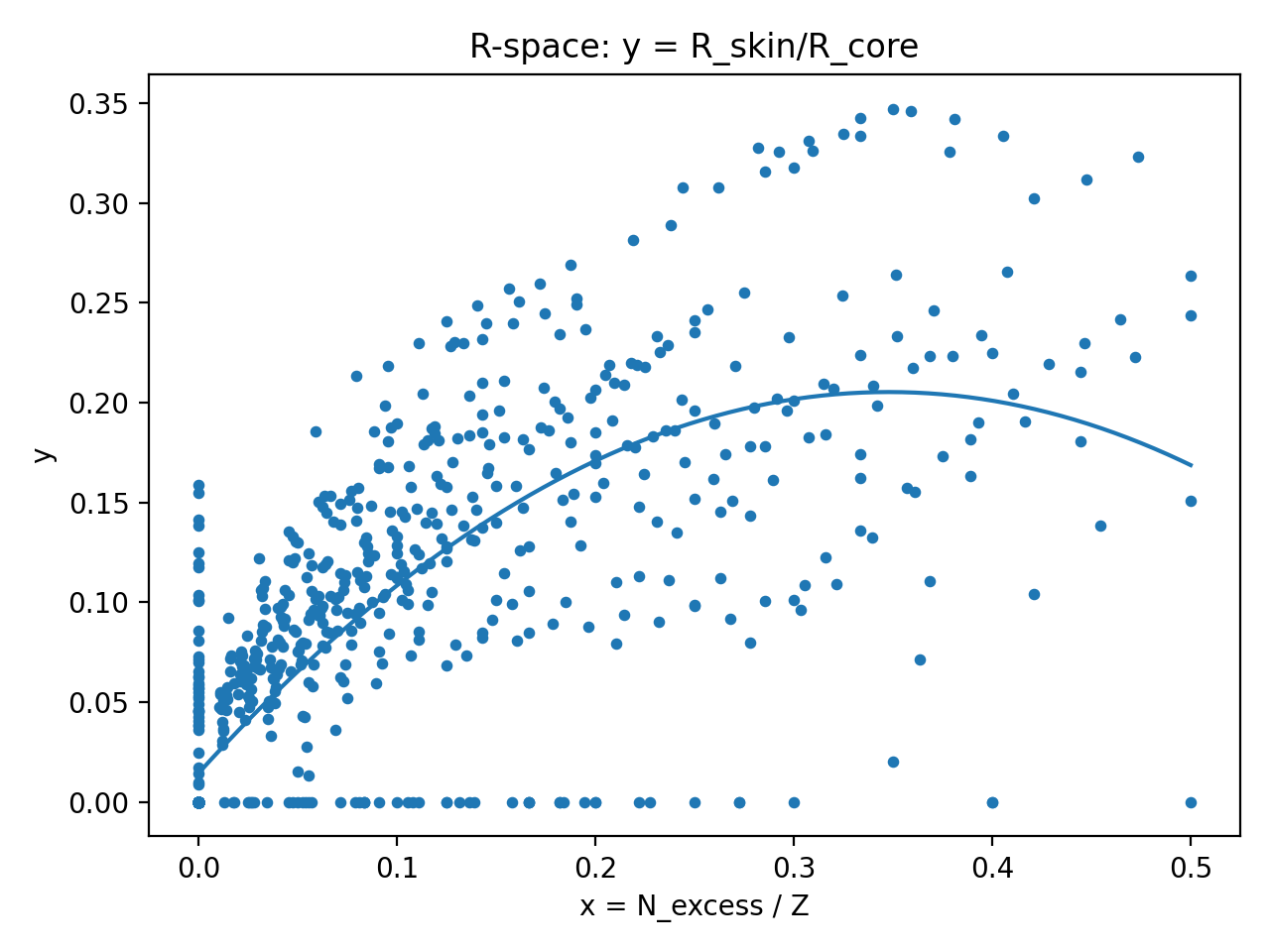}
	\caption{
		Radii-only geometric baseline.
		A skin proxy constructed from experimental charge radii without mass normalization is plotted against $N_{\mathrm{excess}}/Z$.
		While qualitative growth with neutron excess is evident, no cross-element collapse is observed, and substantial element-dependent scatter remains.
	}
	\label{fig:radii_only_baseline}
\end{figure}
While this construction captures the qualitative growth of nuclear size with neutron excess, it does not produce a cross-element collapse. Significant element-dependent scatter remains, reflecting the absence of an explicit mass-based normalization. This behavior is representative of standard radius-systematics plots, where trends are typically analyzed within individual isotopic chains rather than across the full nuclear chart.

\subsection{Droplet--Style Isovector Scaling}

As a stronger standard comparator, we consider a global droplet-style scaling\cite{MyersSwiatecki1969} in which the radii-only skin proxy is fit to a linear function of normalized neutron excess,
\begin{equation}
	\frac{r_{\mathrm{skin}}}{r_{\mathrm{core}}} \approx a \frac{N_{\mathrm{excess}}}{Z} + b,
\end{equation}
with coefficients $a$ and $b$ determined by a least-squares fit over the same validated dataset.

\begin{figure}[t]
	\centering
	\includegraphics[width=0.85\linewidth]{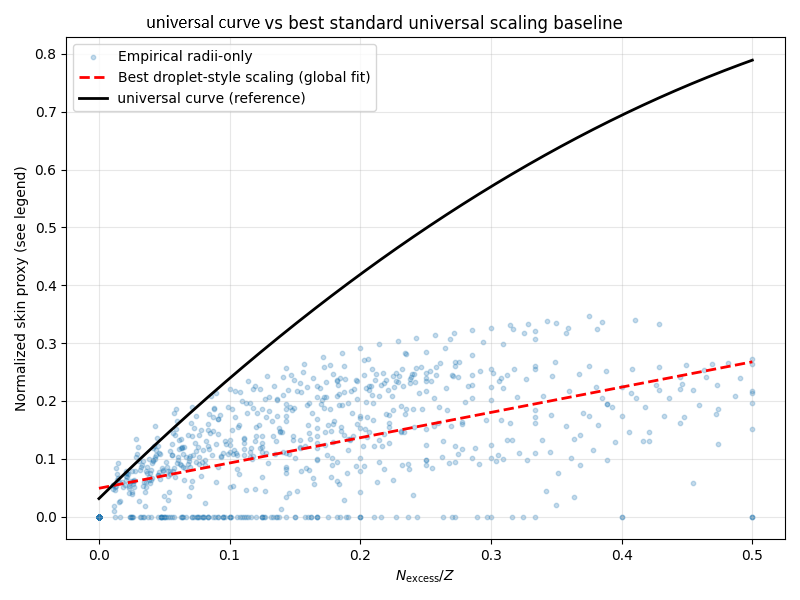}
	\caption{
		Comparison of droplet-style isovector scaling to curvature-normalized scaling.
		The droplet-style baseline (dashed line) is a global linear fit to the radii-only proxy.
		The solid curve shows the derived curvature baseline for reference.
		Despite fitted parameters, the droplet-style baseline does not produce a comparable collapse.}
	\label{fig:droplet_vs_set}
\end{figure}

This form represents the simplest universal isovector trend commonly employed in macroscopic nuclear models. When evaluated against the data, the droplet-style baseline captures a weak global correlation but leaves substantial residual scatter. Quantitatively, it explains approximately $35\%$ of the variance in the radii-only proxy, compared to approximately $88\%$ for the curvature-normalized scaling, despite introducing fitted parameters. The reduced variance collapse of the droplet-style baseline in this global coordinate reflects scope rather than inadequacy: macroscopic and microscopic nuclear models are optimized for local predictive accuracy, not for revealing cross-element dimensionless geometric regularities.
\begin{figure}[t]
	\centering
	\includegraphics[width=0.85\linewidth]{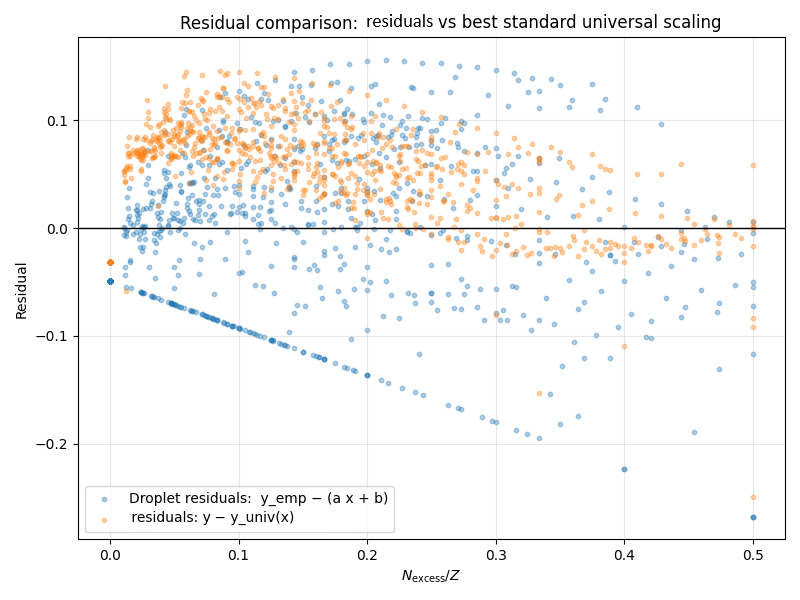}
	\caption{
		Residual comparison between curvature-normalized scaling and the droplet-style baseline.
		Residuals relative to the derived baseline exhibit reduced scatter and minimal structure compared to the droplet-style residuals, indicating improved cross-element organization.
	}
	\label{fig:residuals_comparison}
\end{figure}

\subsection{Relative Performance and Scope}

\begin{table}[ht]
	\centering
	\caption{Comparison of curvature-normalized scaling to standard baselines over the validated domain ($0 \le N_{\mathrm{excess}}/Z \le 0.5$). The derived universal curve uses no fitted parameters in this evaluation (fixed baseline), while the droplet-style baseline uses two globally fitted parameters $(a,b)$ on the radii-only proxy.}
	\label{tab:baseline_comparison}
	\begin{adjustbox}{max width=\textwidth}
	\begin{tabular}{lcccc}
		\toprule
		Method & Parameters & RMSE & $R^2$ & Residual $\sigma$ \\
		\midrule
		Derived universal scaling (fixed $F(x)$) & 0 & 0.0701 & 0.8822 & 0.0493 \\
		Droplet-style baseline ($y=a x + b$ on radii-only proxy) & 2 & 0.0747 & 0.3504 & 0.0747 \\
		\bottomrule
	\end{tabular}
	\end{adjustbox}
	\vspace{0.5em}
	\footnotesize
	\noindent Droplet fit parameters (global): $a=0.4374$, $b=0.0488$.
\end{table}
Table~\ref{tab:baseline_comparison} summarizes the comparative performance of the curvature-normalized scaling and the droplet-style baseline over the validated domain. The curvature-normalized approach achieves a significantly higher variance collapse and reduced residual scatter without introducing fitted parameters or element-dependent adjustments.

While the baseline comparisons above illustrate the qualitative limitations of
radii-only and droplet-style scalings, they do not by themselves establish whether
the improved collapse obtained through curvature normalization reflects a genuinely
new organizing principle or could be reproduced by alternative radius-based
constructions under matched statistical conditions. To address this directly, we
perform an explicit null-hypothesis test comparing radius-based and
curvature-normalized representations constructed from the same experimental charge
radii, using identical core anchors, domain cuts, and baseline model complexity.
That analysis, presented in Appendix~\ref{app:mass_scaling_test}, demonstrates that
curvature normalization yields a statistically significant and generalizable
improvement in global organization beyond what can be achieved using radii-only
coordinates. The results reported here therefore represent not only a descriptive
contrast but are supported by a formal statistical rejection of the null hypothesis
that mass scaling introduces no additional structure.

It is instructive to contrast the curvature-based normalization employed here
with conventional nuclear size scalings commonly used in global systematics.

Conventional nuclear size systematics often employ the empirical scaling relation $R\propto A^{1/3}$\cite{Krane1987}, which removes bulk volume growth under the
assumption of approximately constant nuclear density. While effective for
describing average size trends, this normalization does not eliminate
mass-dependent scaling associated with the underlying energy scale of the
system. As a result, variations arising from neutron excess, shell structure,
and finite-size effects remain intermingled when radii are compared across
elements.

In contrast, one may normalize radii by the reduced Compton wavelength associated with the nuclear mass,
\[
\bar{\lambda}_C=\frac{\hbar}{mc}
\]
\cite{CODATA2018}
which defines a mass-based length scale constructed only from fundamental constants. Expressing measured charge radii as $K_R = r_{\mathrm{exp}}/\bar{\lambda}_C$ yields a dimensionless size variable and removes trivial $1/m$ scaling, allowing nuclei of widely varying mass to be compared on a common geometric footing. When combined with a normalized neutron-excess parameterization, this representation yields a substantially improved collapse of neutron-skin behavior across the nuclear chart.

The improved performance of the curvature-normalized scaling therefore reflects
the use of a normalization that factors out mass-dependent structure more
effectively than volume-based representations, rather than the inclusion of new
dynamical assumptions. This conclusion is further supported by the null-hypothesis 
test reported in Appendix~\ref{app:mass_scaling_test}.

We stress that this comparison does not imply that existing nuclear models are incorrect or incomplete. Rather, the baselines considered here address a different question: how neutron-skin behavior varies within a given isotopic chain or model context. The curvature-normalized scaling instead exposes a dimensionless geometric regularity that is not explicitly targeted by standard formulations.

In subsequent sections, we examine the structured residual behavior that remains after this normalization and show that it encodes physically meaningful finite-size and family-dependent effects.
	
	\section{Residual Structure and Finite--Size Regimes}
\label{sec:residual_regimes}

Although the curvature-normalized scaling described in Sec.~\ref{sec:universal_skin_scaling} produces a strong collapse across the nuclear chart, it is not exact. To assess the nature of the remaining deviations, we examine the residual
\begin{equation}
	\Delta K \equiv \frac{K_{R,\mathrm{skin}}}{K_{R,\mathrm{core}}} - F\!\left(\frac{N_{\mathrm{excess}}}{Z}\right),
\end{equation}
where $F(x)$ denotes the fixed empirical baseline associated with the universal scaling.
\begin{figure}[t]
	\centering
	\includegraphics[width=0.85\linewidth]{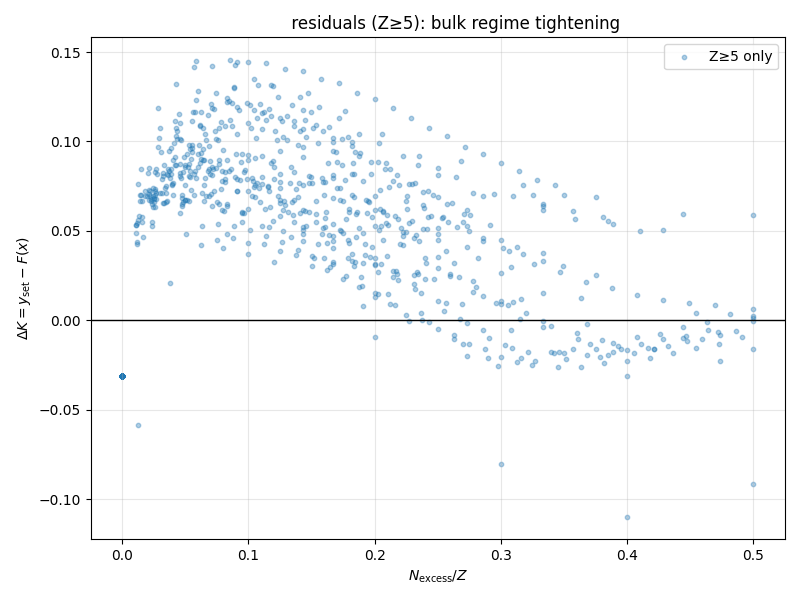}
	\caption{
		Residual neutron-skin curvature $\Delta K$ as a function of normalized neutron excess.
		Three finite-size regimes are observed: an initial skin-formation regime at low $N_{\mathrm{excess}}/Z$, a relaxation regime at intermediate values, and a saturation regime at higher neutron excess.
		The structure indicates physical finite-size effects rather than random scatter.
	}
	\label{fig:residual_regimes}
\end{figure}

The residuals exhibit structured behavior rather than random scatter. When plotted as a function of the normalized neutron excess $N_{\mathrm{excess}}/Z$, three distinct regimes are observed. These regimes appear consistently across many elements and are therefore interpreted as finite-size geometric effects superimposed on the dominant universal trend.

\subsection{Skin--Formation Regime}

At low normalized neutron excess ($N_{\mathrm{excess}}/Z \lesssim 0.05$--$0.08$), the residual variance increases with neutron addition. In this regime, the neutron skin is just beginning to form, and small changes in neutron number can produce relatively large fractional changes in surface geometry. As a result, isotopic behavior is more sensitive to shell structure, pairing, and local geometric rearrangements.

This increase in residual scatter is expected in any description where the skin is not yet a well-defined geometric feature. The universal scaling should therefore be understood as an asymptotic description rather than as an exact law at vanishing neutron excess.

\subsection{Relaxation Regime}

At intermediate neutron excess ($\sim 0.08 \lesssim N_{\mathrm{excess}}/Z \lesssim 0.3$), the mean residual exhibits a gradual, approximately linear decrease toward zero, and the overall scatter narrows. This behavior is consistent with convergence toward a bulk-like geometric regime as the neutron skin becomes established.

In this relaxation regime, incremental neutrons contribute more uniformly to the surface, and element-to-element differences are increasingly suppressed by the mass-based normalization. The observed trend suggests a finite-size relaxation process toward the asymptotic curvature configuration represented by the universal scaling.

\subsection{Saturation Regime}

At higher neutron excess ($N_{\mathrm{excess}}/Z \gtrsim 0.3$), the residuals flatten and the variance stabilizes. In this regime, additional neutrons primarily populate an already-formed surface, and further changes in neutron excess do not significantly alter the normalized curvature ratio.

The existence of this saturation regime supports the interpretation that the universal curve represents a limiting surface-geometry behavior, with deviations at lower neutron excess reflecting transitional finite-size effects rather than failure of the scaling.

\subsection{Interpretation}

Taken together, these three regimes indicate that the curvature-normalized scaling captures the dominant geometric behavior of neutron skins, while the residuals encode secondary structure associated with finite size, pairing, and shell effects. Importantly, the residual patterns are systematic and reproducible across the dataset, reinforcing the conclusion that they are physical rather than statistical in origin.
An exploratory overlay of evaluated nuclear decay modes in the same normalized coordinate is presented in Appendix~\ref{app:decay_mode_overlay} as a descriptive diagnostic aligning with the apparent physical correlation interpretation.

In the following sections, we examine how these residuals depend on nuclear domain and family classification, beginning with the treatment of very light nuclei and the limits of applicability of a bulk skin description.
	\section{Light-Nucleus Domain of Validity}
\label{sec:light_nucleus_validity}

The curvature-normalized scaling presented above is intended to describe bulk neutron-skin geometry. It is therefore important to identify regimes in which the concept of a neutron skin is not physically meaningful or is expected to behave qualitatively differently.
\begin{figure}[t]
	\centering
	\includegraphics[width=0.85\linewidth]{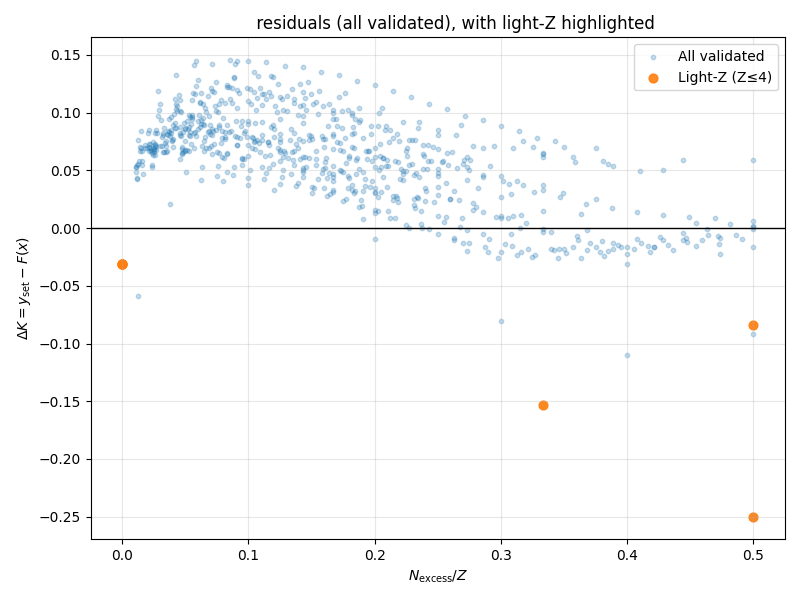}
	\caption{
		Residuals with light nuclei ($Z \le 4$) highlighted.
		Very light nuclei occupy a few-body and clustering-dominated regime and contribute disproportionately to extreme residuals.
		Their behavior is distinct from bulk neutron-skin geometry.
	}
	\label{fig:lightZ_highlight}
\end{figure}
Nuclei with very small proton number ($Z \le 4$) occupy a few-body and clustering-dominated regime. In these systems, the neutron--proton configuration is governed by strong correlations and discrete cluster structure rather than by a smooth core-plus-surface geometry. In the extreme case of hydrogen ($Z=1$), no neutron skin exists at all, and the core--skin decomposition becomes undefined. Even for helium and lithium isotopes, the notion of a continuous neutron surface is not appropriate in the same sense as for medium and heavy nuclei.

Table~\ref{tab:lightZ_robustness}. Consistent with this expectation, nuclei with $Z \le 4$ exhibit systematically larger curvature residuals and dominate the extreme tails of the residual distribution. When these nuclei are excluded, the residual scatter tightens and the most extreme deviations collapse, while the overall universal trend remains unchanged. Quantitatively, excluding the $Z \le 4$ subset reduces the residual standard deviation and improves the robustness of the collapse without altering its functional form.
\begin{table}[t]
	\centering
	\caption{Light-$Z$ robustness check for the universal scaling. Excluding the few-body regime ($Z \le 4$) tightens the residual distribution and collapses the extreme negative tail, while leaving the overall trend unchanged.}
	\label{tab:lightZ_robustness}
	\begin{tabular}{lrrrrrr}
		\toprule
		Subset & Rows & $Z$ chains & RMSE & $R^2$ & Residual $\sigma$ & $\min(\Delta K)$ \\
		\midrule
		All validated & 826 & 88 & 0.070118 & 0.882202 & 0.049312 & $-0.249826$ \\
		$Z \ge 5$ (exclude $Z \le 4$) & 819 & 84 & 0.069573 & 0.882619 & 0.047300 & $-0.109784$ \\
		$Z \le 4$ only & 7 & 4 & 0.117570 & 0.834892 & 0.078807 & $-0.249826$ \\
		\bottomrule
	\end{tabular}
	
	\vspace{0.5em}
	\footnotesize
	\noindent Here $\Delta K = \big(K_{R,\mathrm{skin}}/K_{R,\mathrm{core}}\big) - F(N_{\mathrm{excess}}/Z)$.
\end{table}

We emphasize that this exclusion does not constitute tuning or data selection. Rather, it reflects a physically motivated domain-of-validity criterion: the universal scaling describes bulk neutron-skin geometry and is not expected to apply in few-body or cluster-dominated systems. The light-nucleus subset is therefore treated as a distinct geometric regime and is not used to characterize the asymptotic behavior of neutron skins across the nuclear chart.

All subsequent analyses of residual structure and family-dependent behavior focus on nuclei with $Z \ge 5$, for which the core--skin decomposition is well-defined and the neutron-skin concept is physically meaningful.
	\section{Family--Conditioned Residual Structure}
\label{sec:family_conditioned_structure}

The universal scaling described in Secs.~\ref{sec:universal_skin_scaling}--\ref{sec:residual_regimes} captures the dominant behavior of neutron-skin curvature across the nuclear chart. However, the residuals $\Delta K$ exhibit additional structure that is not random. To investigate whether this structure correlates with known classifications of nuclei, we examine the residuals after stratifying the data by nuclear and chemical families.

\subsection{Method of Family Stratification}

Two complementary family classifications are considered. First, nuclei are grouped by conventional periodic-table families (e.g., alkali metals, alkaline earth metals, transition metals, lanthanides, actinides, noble gases). These families are traditionally defined by electronic structure but serve here purely as empirical labels. Second, as a result of investigating individual isotopic chain quadratics while developing the global trend, more specific nuclear curvature families are proposed as an independent classification in Appendix \ref{app:family_branching} as to not distract from the focus of this paper.

In both cases, no refitting or curve adjustment is performed. The same universal baseline $F(x)$ is used throughout, and family structure is assessed solely through the statistical properties of the residuals.

\subsection{Residual Tightening Within Families}
\begin{figure}[t]
	\centering
	\includegraphics[width=0.9\linewidth]{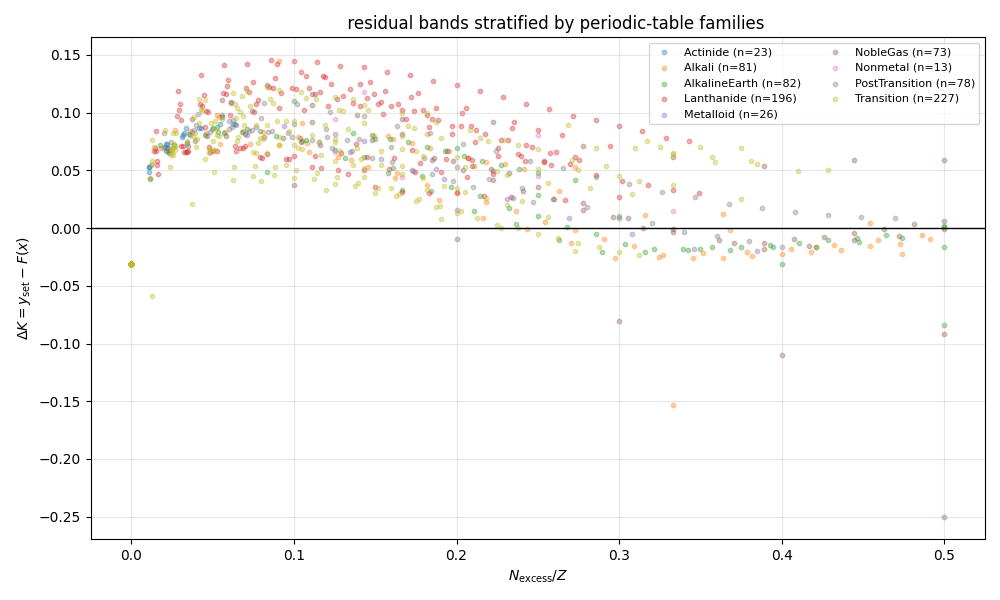}
	\caption{
		Residual neutron-skin curvature stratified by periodic-table family.
		Several heavy-element families form tight sub-manifolds around the universal baseline, while families dominated by light or closed-shell nuclei exhibit broader scatter.
	}
	\label{fig:family_bands}
\end{figure}

When residuals are examined within individual families, several groups exhibit a marked reduction in scatter relative to the global distribution. In particular, transition metals, post-transition metals, and lanthanides form noticeably tighter sub-manifolds around the universal curve, with residual standard deviations reduced by approximately 10--20\% compared to the full dataset. Alkaline earth metals and actinides show moderate tightening, while families dominated by light or closed-shell nuclei (e.g., noble gases, halogens, nonmetals) exhibit broader residual distributions.
\begin{table}[t]
	\centering
	\caption{Residual statistics by conventional periodic-table family (empirical labels). Several heavy families (e.g., transition metals, lanthanides, post-transition metals) form tighter sub-manifolds around the universal curve, while families dominated by light or closed-shell nuclei show broader residual distributions.}
	\label{tab:periodic_family_tightening}
	\begin{tabular}{lrrrr}
		\toprule
		Family & $n$ & Mean $\Delta K$ & Residual $\sigma$ & $\sigma/\sigma_{\mathrm{global}}$ \\
		\midrule
		Post-transition metals & 78  & 0.051844 & 0.039578 & 0.799636 \\
		Lanthanides           & 196 & 0.078729 & 0.042443 & 0.857520 \\
		Transition metals      & 227 & 0.051633 & 0.042648 & 0.861668 \\
		Alkaline earth metals  & 82  & 0.028293 & 0.044114 & 0.891285 \\
		Actinides              & 23  & 0.053108 & 0.046996 & 0.949525 \\
		Metalloids             & 26  & 0.032028 & 0.047878 & 0.967343 \\
		Alkali metals          & 81  & 0.026016 & 0.049119 & 0.992409 \\
		Nonmetals              & 13  & 0.021322 & 0.056051 & 1.132463 \\
		Halogens               & 6   & 0.005573 & 0.056942 & 1.150459 \\
		Noble gases            & 73  & 0.029671 & 0.060895 & 1.230340 \\
		\bottomrule
	\end{tabular}
	
	\vspace{0.5em}
	\footnotesize
	\noindent $\sigma_{\mathrm{global}}$ is the residual standard deviation over all family-labeled nuclei used in this table. Small-$n$ families (e.g., halogens) carry larger uncertainty and are reported for completeness.
\end{table}

Table~\ref{tab:periodic_family_tightening} This behavior is consistent with known nuclear-structure trends. Families characterized by heavier nuclei and extended isotopic chains tend to display smoother geometric evolution and therefore tighter residual clustering. In contrast, families associated with closed shells, few available isotopes, or pronounced shell effects naturally show greater scatter when analyzed in a universal framework.
\begin{figure}[t]
	\centering
	\includegraphics[width=0.9\linewidth]{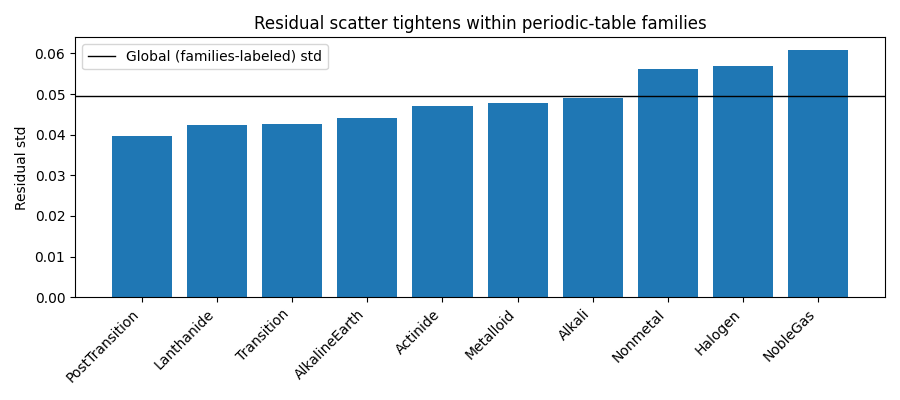}
	\caption{
		Residual standard deviation by periodic-table family.
		Values are normalized to the global residual scatter.
		Several families show a $10$--$20\%$ reduction in residual variance, indicating tighter geometric organization within those groups.
	}
	\label{fig:family_std}
\end{figure}

\subsection{Interpretation}

The emergence of tighter residual bands upon family stratification indicates that the remaining deviations from the universal scaling encode secondary geometric constraints rather than noise. Importantly, this effect is revealed without introducing additional parameters or modifying the baseline curve. The family-conditioned structure therefore reflects genuine physical organization layered on top of the dominant universal behavior.

We stress that the use of periodic-table families does not imply that electronic structure causes the observed nuclear behavior. Rather, these families act as convenient empirical groupings that correlate with underlying nuclear properties such as mass range, shell structure, and deformation tendencies. The fact that independent family classifications produce coherent residual patterns supports the interpretation that neutron-skin curvature contains structured information beyond its leading-order scaling.

In the following discussion, we place these results in the context of existing nuclear-structure studies and outline how the family-conditioned residuals motivate further investigation without altering the empirical foundation established here.
	
	\section{Discussion}
\label{sec:discussion}
From a physical standpoint, the significance of the present scaling lies in its ability to organize a charge-radius--derived neutron-excess surface proxy across elements using a single dimensionless coordinate.
Traditional neutron-skin thickness and related observables are commonly discussed in connection with symmetry-energy systematics, neutron radii of heavy nuclei (e.g.\ $^{208}$Pb), and surface properties extracted within model-dependent frameworks\cite{Horowitz2014,PREXII2021,CREX2022}.
The curvature-normalized representation introduced here is not an extraction of $r_n$ (and thus not a direct determination of $\Delta r_{np}$); rather, it provides a compact, model-agnostic reference baseline for a radii-derived proxy against which such observables can be compared or stratified across the nuclear chart.

The results presented in this work demonstrate that neutron-skin behavior across the nuclear chart admits a compact, dimensionless description when experimental charge radii are expressed in a mass-normalized curvature framework. Using only evaluated nuclear radii, masses, and fundamental constants, we identify a universal scaling of neutron-skin curvature as a function of normalized neutron excess that collapses data from hundreds of nuclei onto a single curve.

\subsection{Interpretation of the Universal Scaling}

The emergence of a universal collapse does not imply that all nuclei share identical internal structure. Rather, it indicates that the dominant geometric contribution to neutron-skin growth is controlled by a small number of dimensionless variables once trivial mass dependence is removed. In this sense, the scaling should be understood as an asymptotic geometric regularity, analogous to other dimensionless trends that emerge only after appropriate normalization.

The fact that this collapse is achieved without fitted parameters or element-dependent adjustments suggests that the mass-based normalization plays a central role in exposing the underlying structure. Importantly, the normalization does not introduce new dynamics; it reorganizes existing experimental information in a way that reveals previously obscured regularity.

\subsection{Residual Structure as Physical Information}

The scaling is not exact, and the residuals exhibit systematic structure rather than random scatter. As shown in Sec.~\ref{sec:residual_regimes}, the residual behavior organizes naturally into three regimes corresponding to skin formation, relaxation toward a bulk geometry, and saturation at larger neutron excess. These regimes are consistent with finite-size effects and the gradual establishment of a well-defined neutron surface.

Further stratification by nuclear and chemical families reveals that several groups form tighter sub-manifolds around the universal trend. This observation indicates that residual structure encodes secondary geometric constraints associated with shell structure, deformation, and mass range. The fact that these patterns emerge without refitting reinforces the conclusion that the residuals are physically meaningful rather than artifacts of noise or overfitting.

\subsection{Domain of Applicability}

The analysis also clarifies the domain of applicability of the universal scaling. Very light nuclei ($Z \le 4$) occupy a few-body regime in which the concept of a bulk neutron skin is not physically appropriate. Their exclusion sharpens the collapse and reduces extreme residuals without altering the overall trend, confirming that the scaling pertains to bulk nuclear geometry rather than to cluster-dominated systems.

This domain-of-validity criterion is physically motivated and aligns with standard nuclear-structure understanding. It should therefore be viewed as a feature of the analysis rather than as a limitation.

\subsection{Relation to Existing Nuclear Models}

The present results do not contradict existing nuclear models or phenomenological descriptions. Rather, they address a complementary question. Standard models are optimized to reproduce detailed properties of individual nuclei or isotopic chains, often with element-specific parameters. The curvature-normalized scaling instead reveals a cross-element geometric regularity that is not explicitly targeted by those approaches.

In this sense, the scaling should be viewed as an additional organizing principle that coexists with established models. It does not replace shell-model, energy-density-functional, or droplet-model descriptions, but highlights a dimensionless structure that becomes apparent only when nuclear sizes are compared on a common mass-dependent scale.

\subsection{Implications and Outlook}

The existence of a universal neutron-skin scaling has several implications. First, it provides a compact empirical framework for comparing neutron-skin behavior across elements and isotopic chains. Second, the structured residuals offer a quantitative way to identify regions where secondary geometric effects—such as pairing and deformation—become important. Finally, the dimensionless nature of the scaling facilitates comparison with other observables that depend on nuclear geometry.
Several global response observables, such as dipole polarizability, are known to vary only weakly across isotopic chains when expressed in raw form. This behavior is consistent with geometric rebalancing: as neutron skins grow and nuclear size increases, global response characteristics remain approximately stable after bulk size effects are accounted for. The present work isolates the underlying geometric evolution that precedes such rebalancing. A microscopic interpretation of the curvature normalization and its relation to nuclear dynamics is beyond the scope of this paper and is deferred to future work and is not required for the empirical validity of the results reported here.

While the present work focuses exclusively on nuclear radii, the same curvature variables can be evaluated independently of spectroscopic or decay data. This separation allows subsequent studies to test whether the geometric organization identified here propagates into other domains without circularity. Such extensions are the subject of ongoing work and are not required for the validity of the empirical results reported in this paper. A detailed theoretical interpretation of the curvature normalization and its relation to wave–mass duality, projection limits, and observer structure is developed separately and is not required for the empirical validity of the present results.
\subsection{Relation to Nuclear Decay Systematics}

While the present work focuses on neutron-skin geometry inferred from charge radii,
it is natural to ask whether the same normalized coordinate organizing neutron-skin behavior
also provides a useful perspective on nuclear stability.
As an exploratory diagnostic, evaluated dominant decay modes from the NuDat/ENSDF databases
were overlaid on the normalized neutron-excess coordinate using the same element-specific
core anchors as the radii analysis.

As shown in Appendix~\ref{app:decay_mode_overlay}, decay channels are not uniformly distributed
in this coordinate.
Proton emission and EC/$\beta^+$ dominate at negative values of $N_{\mathrm{excess}}/Z$,
$\alpha$ decay appears primarily for heavy nuclei near $N_{\mathrm{excess}}/Z \approx 0$,
and $\beta^-$ decay becomes increasingly dominant for positive values of $N_{\mathrm{excess}}/Z$.
Spontaneous fission is largely absent from the radii-validated domain and appears only
for very heavy nuclei outside the range of long-lived isotopes with measured charge radii.

This association is reported here solely as an empirical observation.
No claim is made that neutron-skin curvature determines decay pathways,
nor that decay modes can be predicted from the curvature framework alone.
Rather, the overlay demonstrates that the same normalized coordinate that organizes
neutron-skin geometry also provides a compact way to visualize decay-channel systematics,
motivating more detailed future studies of stability boundaries and decay dynamics.

	\section{Conclusion}
\label{sec:conclusion}

In this work we have presented an empirical analysis of neutron-skin geometry across the nuclear chart using a mass-normalized curvature framework derived directly from experimental charge radii and evaluated nuclear masses. By expressing nuclear size in terms of a dimensionless curvature ratio and isolating the contribution of excess neutrons through a core--skin decomposition, we identify a universal scaling of neutron-skin curvature as a function of normalized neutron excess.

This scaling collapses data from more than 800 nuclei across 88 elements onto a single curve without introducing fitted parameters or element-dependent adjustments. When compared to standard baselines, including radii-only constructions and droplet-style isovector scaling, the curvature-normalized representation achieves a substantially stronger variance collapse and reduced residual scatter. These results demonstrate that an appropriate mass-based normalization exposes cross-element geometric regularity that is not apparent in conventional representations.

The remaining deviations from the universal trend are structured rather than random. Analysis of the residuals reveals finite-size regimes associated with skin formation, relaxation toward a bulk surface geometry, and saturation at larger neutron excess. Additional stratification by nuclear and chemical families further tightens the residual distributions for several groups, indicating the presence of secondary geometric constraints layered on top of the dominant universal behavior. Very light nuclei are shown to occupy a distinct few-body regime and are appropriately excluded from characterization of bulk neutron-skin geometry.

All results reported here arise directly from experimental data and standard physical constants. No new interaction terms are introduced, and no existing nuclear models are modified or displaced. Instead, the present analysis provides an additional, dimensionless organizational framework that complements established descriptions by revealing a previously obscured regularity across the nuclear chart.

The empirical scaling and structured residual behavior identified in this work establish a robust foundation for further investigation of nuclear geometry and its connections to other observables. Extensions to atomic spectroscopy and broader interpretive frameworks can be pursued independently, but are not required for the validity of the results presented here.

	\appendix
	\section{Data Sources and Processing}
\label{app:data_sources}

All results presented in this work are derived exclusively from evaluated, publicly available nuclear datasets. No proprietary data or unpublished measurements are used.

\subsection{Charge Radii}

Experimental nuclear charge radii $r_{\mathrm{exp}}$ are taken from the evaluated compilation of Angeli and Marinova~\cite{AngeliMarinova2013}. This dataset aggregates measurements from electron scattering, muonic atom spectroscopy, and optical isotope-shift studies, and represents the standard reference for nuclear charge radii across the chart.
\begin{table}[t]
	\centering
	\caption{Primary datasets used in this paper and their roles in the analysis pipeline.}
	\label{tab:dataset_provenance}
	\begin{adjustbox}{max width=\textwidth}
	\begin{tabular}{lll}
		\toprule
		Dataset & Quantity used & Role in analysis \\
		\midrule
		Angeli \& Marinova (2013)~\cite{AngeliMarinova2013} & $r_{\mathrm{exp}}$ (charge radii) & Observed nuclear size input \\
		AME2020~\cite{AME2020} & Nuclear mass $m$ & Mass-based normalization $\bar{\lambda}_C=\hbar/(mc)$ \\
		CODATA 2018~\cite{CODATA2018} & $c,h,\hbar$ & Physical constants for normalization \\
		\bottomrule
	\end{tabular}
	\end{adjustbox}
\end{table}
When multiple measurements are available for a given isotope, the evaluated value reported in the compilation is used directly. No smoothing, interpolation, or extrapolation is applied.

\subsection{Nuclear Masses}

Nuclear masses are taken from the Atomic Mass Evaluation (AME2020)~\cite{AME2020}. Atomic masses are converted to nuclear masses using standard electron-binding corrections where required. These masses are used solely to compute the mass-based normalization scale defined in Sec.~\ref{sec:Mass-Based Radius Normalization}.
\subsection{Physical Constants}

Fundamental constants ($c$, $h$, $\hbar$) are adopted from the CODATA 2018 recommended values~\cite{CODATA2018}. Constant uncertainties are negligible compared to experimental radius uncertainties and do not affect any qualitative or quantitative conclusions.

\subsection{Dataset Filtering}

Only isotopes satisfying the following criteria are included in the validated dataset:
\begin{itemize}[leftmargin=*]
	\item A measured and evaluated charge radius is available.
	\item A corresponding evaluated nuclear mass is available.
	\item The isotope is long-lived on experimental timescales.
	\item The normalized neutron excess satisfies $0 \le N_{\mathrm{excess}}/Z \le 0.5$.
\end{itemize}

After filtering, the validated dataset comprises more than 800 nuclei spanning 88 elements.

	\section{Statistical Methods}
\label{app:statistical_methods}

This work employs simple, transparent statistical measures to quantify the quality of data collapse and baseline comparisons. No advanced fitting or machine-learning techniques are used.

\subsection{Baseline Comparisons}

For each baseline construction, goodness of collapse is evaluated using:
\begin{itemize}[leftmargin=*]
	\item The coefficient of determination ($R^2$),
	\item The root-mean-square error (RMSE),
	\item The standard deviation of residuals.
\end{itemize}

These metrics are computed over the same validated dataset for all baselines to ensure consistency.

\subsection{Residual Definition}

Residuals are defined relative to the fixed empirical baseline $F(x)$ associated with the universal neutron-skin scaling:
\begin{equation}
	\Delta K = \frac{K_{R,\mathrm{skin}}}{K_{R,\mathrm{core}}} - F\!\left(\frac{N_{\mathrm{excess}}}{Z}\right).
\end{equation}

No refitting is performed when evaluating residuals for subsets of the data.

\subsection{Family-Conditioned Statistics}

To assess family-conditioned structure, residual statistics are computed independently for each family using the same baseline. The residual standard deviation within each family is compared to the global residual standard deviation as a measure of relative tightening or broadening.

Families with very small sample sizes are reported but not emphasized, as statistical uncertainty is larger in those cases.

	\section{Validation and Scope of Mass--Scaled Nuclear Geometry}
\label{app:mass_scaling_test}

\subsection{Motivation and null hypothesis}

The curvature-normalized scaling introduced in the main text expresses experimental
charge radii relative to the mass-based length scale
\(
\bar{\lambda}_C = \hbar/(mc).
\)
A natural concern is whether the resulting global organization reflects a genuinely
new geometric regularity, or whether comparable behavior could be obtained using
radius-based constructions alone.

To address this explicitly, we perform a null-hypothesis test using only experimental
nuclear charge radii and evaluated nuclear masses, without invoking spectroscopy,
family labels, or residual conditioning.

\paragraph{Null hypothesis (H$_0$).}
After applying the same core--skin decomposition and the same neutron-excess coordinate
\(x = N_{\mathrm{excess}}/Z\), curvature normalization provides no improvement in global
organization beyond radius-based coordinates.

\paragraph{Alternative hypothesis (H$_1$).}
Curvature normalization removes mass-linked scaling and yields a statistically
significant improvement in the global organization of neutron-skin geometry across
the nuclear chart.

\subsection{Coordinate constructions}

Both representations are constructed from the same experimental charge radii and use
identical element-specific core anchors.

\paragraph{Radius--based representation.}
For each isotope,
\begin{equation}
	R_{\mathrm{skin}} \equiv \sqrt{R_{\mathrm{exp}}^2 - R_{\mathrm{core}}^2},
	\qquad
	y_R \equiv \frac{R_{\mathrm{skin}}}{R_{\mathrm{core}}}.
\end{equation}
This representation measures surface growth relative to a chain-specific geometric
reference.

\paragraph{Curvature--normalized representation.}
Using the mass-based length scale \(\bar{\lambda}_C\),
\begin{equation}
	K_R \equiv \frac{R_{\mathrm{exp}}}{\bar{\lambda}_C},
	\qquad
	K_{R,\mathrm{skin}} \equiv \sqrt{K_R^2 - K_{R,\mathrm{core}}^2},
	\qquad
	y_K \equiv \frac{K_{R,\mathrm{skin}}}{K_{R,\mathrm{core}}}.
\end{equation}
This representation measures surface growth after removing mass-linked scaling.

In both cases, the same neutron-excess coordinate
\(
x = N_{\mathrm{excess}}/Z
\)
is employed.

\subsection{Statistical methodology}

The comparison is restricted to the validated domain
\(Z \ge 5\) and \(0 \le x \le 0.5\).
For both representations, the same quadratic baseline model
\(
y(x) = ax^2 + bx + c
\)
is fitted using ordinary least squares.

Goodness of organization is quantified using:
(i) the coefficient of determination \(R^2\),
(ii) the root-mean-square error (RMSE),
and (iii) the normalized RMSE (RMSE divided by the standard deviation of \(y\)).

To test generalization across the nuclear chart, an element-wise leave-one-out
cross-validation is performed: all isotopes of a given element are excluded from the
fit, and the baseline inferred from the remaining elements is evaluated on the held-out
element.

Finally, a nonparametric bootstrap is used to estimate the sampling distribution of the
difference in normalized error between the curvature-normalized and radius-based
representations.

\subsection{Results}
\begin{figure}[t]
	\centering
	\includegraphics[width=0.85\linewidth]{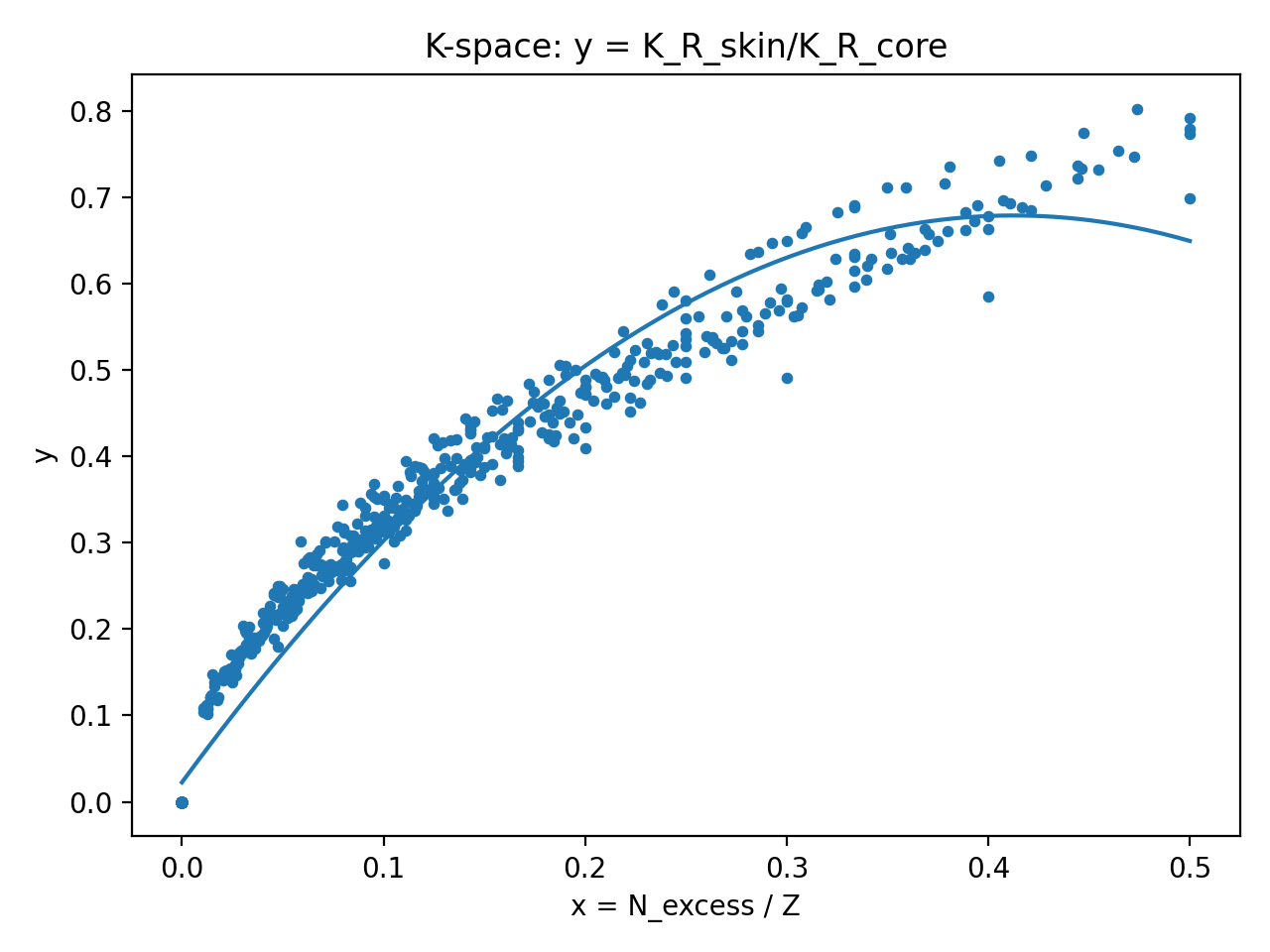}
	\caption{
		Global organization of neutron-skin growth in curvature-normalized coordinates,
		$y = K_{R,\mathrm{skin}}/K_{R,\mathrm{core}}$, as a function of
		$x = N_{\mathrm{excess}}/Z$.
		A fixed quadratic baseline is shown for reference.
	}
	\label{fig:rk_k_space}
\end{figure}
\begin{figure}[t]
	\centering
	\includegraphics[width=0.85\linewidth]{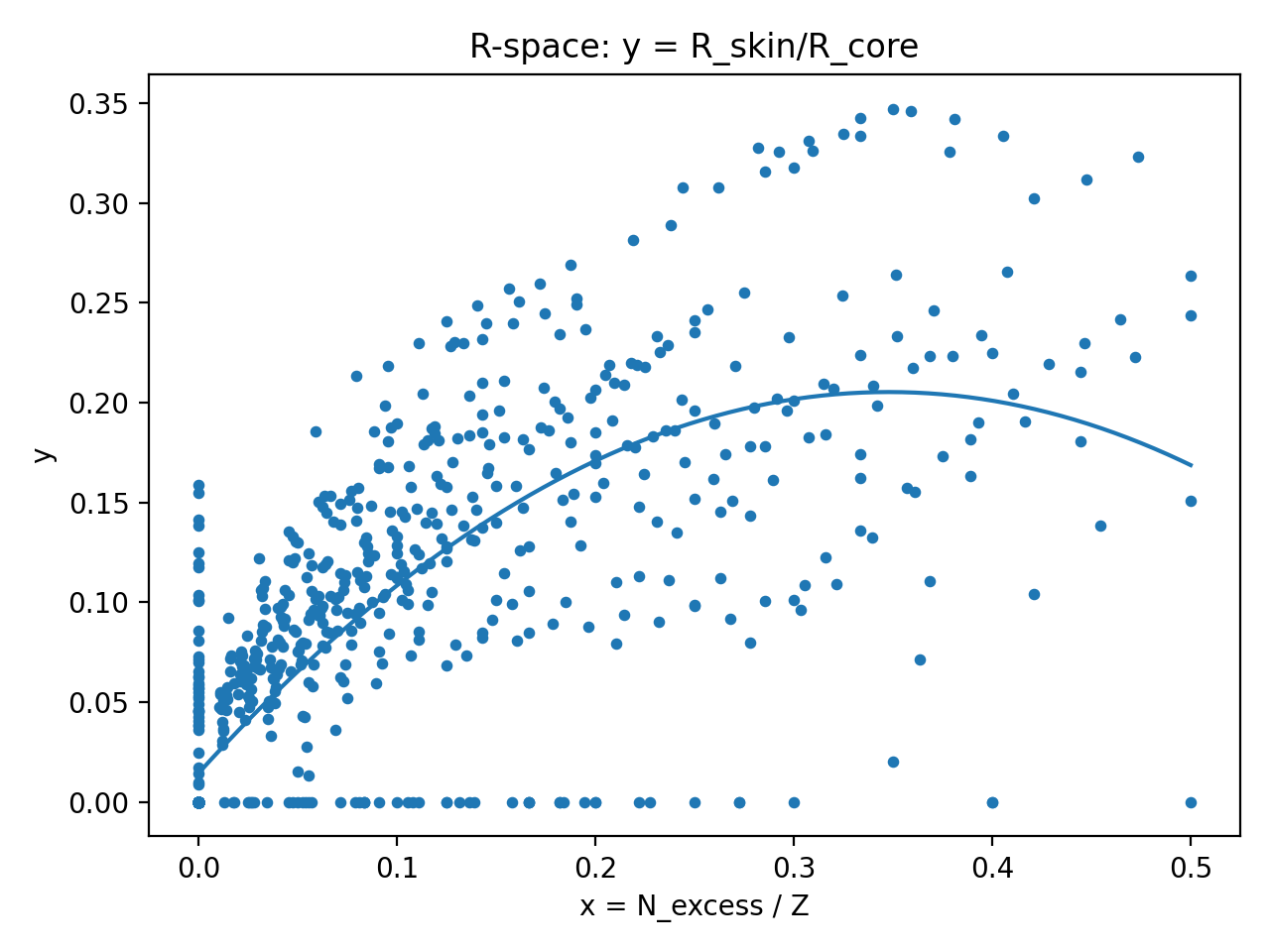}
	\caption{
		Radius-based representation of neutron-skin growth,
		$y = R_{\mathrm{skin}}/R_{\mathrm{core}}$, plotted against
		$x = N_{\mathrm{excess}}/Z$ using the same core anchors and domain.
		While qualitative growth is evident, substantial element-dependent scatter remains.
	}
	\label{fig:rk_r_space}
\end{figure}
\begin{figure}[t]
	\centering
	\includegraphics[width=0.75\linewidth]{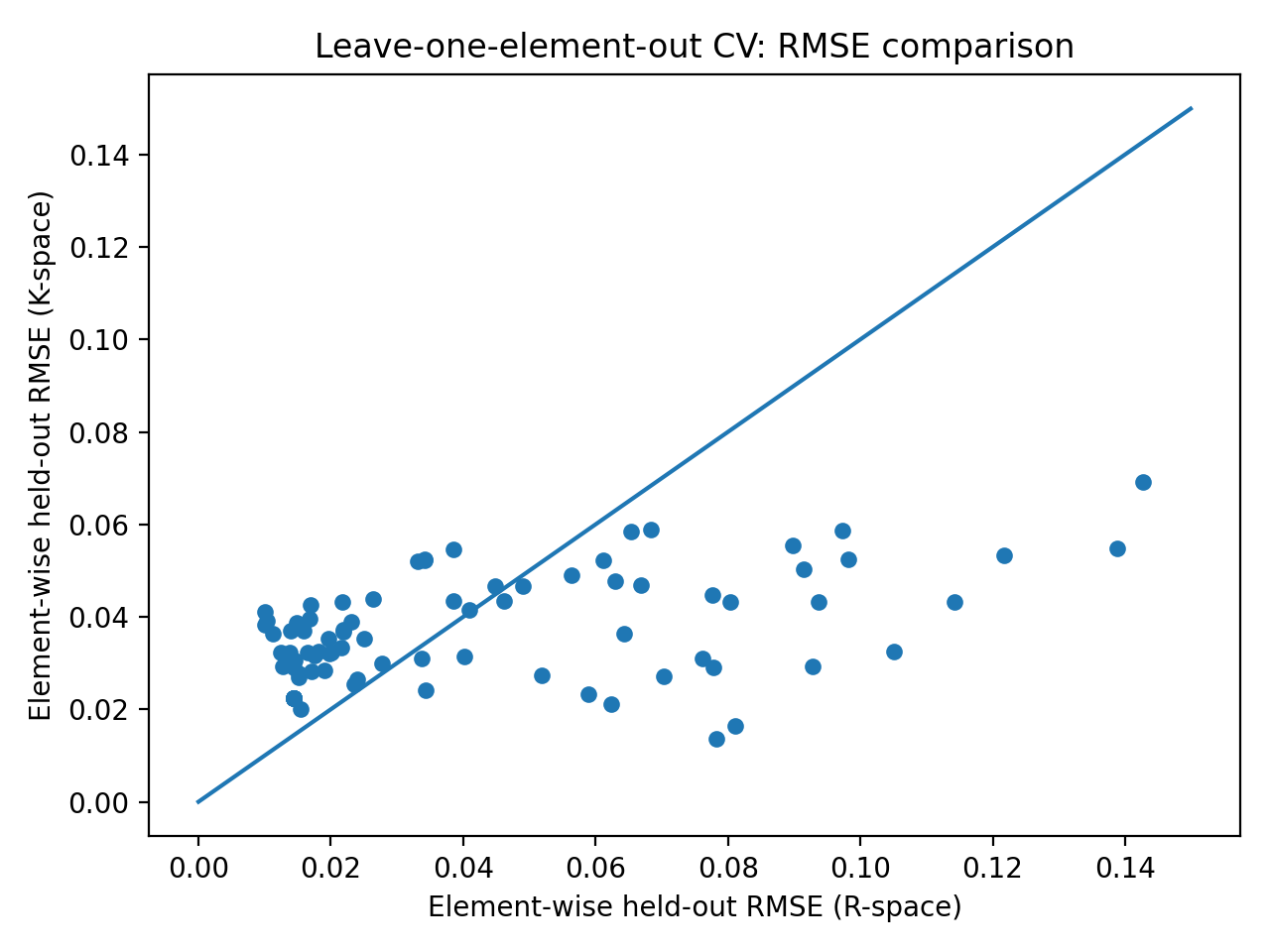}
	\caption{
		Element-wise leave-one-out cross-validation comparing radius-based and
		curvature-normalized representations.
		Each point corresponds to a single element held out during baseline fitting.
		Points lying below the diagonal indicate improved generalization in curvature space.
	}
	\label{fig:rk_cv_rmse_comparison}
\end{figure}

Figures~\ref{fig:rk_k_space} and~\ref{fig:rk_r_space} present the global organization of
neutron-skin growth in curvature-normalized and radius-based representations,
respectively, using the same experimental charge radii, core anchors, and validated
domain.
In both cases, neutron-skin growth increases monotonically with normalized neutron
excess within individual isotope chains.
However, when isotope chains are combined across the nuclear chart, the two
representations differ markedly in their degree of global organization.

In curvature-normalized coordinates
(Fig.~\ref{fig:rk_k_space}), the data collapse tightly onto a single smooth backbone,
with limited scatter about a fixed quadratic baseline.
In contrast, the radius-based representation
(Fig.~\ref{fig:rk_r_space}) exhibits substantially broader dispersion, with pronounced
element-dependent spread that persists despite identical core--skin decomposition and
normalization by the core radius.
This visual contrast indicates that radius-based coordinates preserve local chain
structure but do not support a stable cross-element reference.

Quantitatively, the curvature-normalized representation achieves a coefficient of
determination \(R^2 \approx 0.97\), compared to \(R^2 \approx 0.63\) for the radius-based
representation, together with a reduction in normalized residual scatter by more than a
factor of three under matched quadratic baselines.

The difference between the two representations is further highlighted by the
element-wise leave-one-out cross-validation shown in
Fig.~\ref{fig:rk_cv_rmse_comparison}.
Each point corresponds to an entire element held out during baseline fitting.
The majority of points lie below the diagonal, indicating that curvature-normalized
coordinates systematically yield lower held-out error than radius-based coordinates.
Weighted cross-validation metrics confirm that this improvement persists under element
mixing and does not arise from overfitting to specific isotope chains.

Finally, bootstrap resampling of the difference in normalized error between the two
representations rejects the null hypothesis that curvature normalization provides no
additional organization.
At the sampling resolution employed, the probability that the radius-based
representation outperforms the curvature-normalized representation is consistent with
zero.

\subsection{Interpretation and scope}

This experiment demonstrates that mass scaling has a measurable and statistically
significant effect on the organization of neutron-skin geometry derived from
experimental charge radii.
Radius-based coordinates preserve local structure within individual isotope chains but
do not support a stable global reference when disparate nuclear configurations are
combined.
Curvature normalization removes mass-linked scaling, isolating collective surface
growth and yielding a substantially more coherent global organization.

No claim is made here regarding the physical origin of the mass-based normalization or
its dynamical interpretation.
The result establishes only that curvature normalization provides information beyond
radius-only coordinates when organizing experimental nuclear charge radii.
\subsection{Proton--direction test: isotones}

To assess whether the benefits of mass scaling extend beyond neutron-driven surface
growth, an orthogonal test is performed by examining isotone sequences, in which the
neutron number $N$ is held fixed while the proton number $Z$ varies.
This provides a proton-direction analog to isotopic chains and allows the role of
Coulomb-driven geometry to be evaluated under the same normalization framework.

For selected neutron numbers with broad experimental coverage (e.g., $N=82$, $N=58$,
$N=90$), nuclei are compared across increasing $Z$ using isotone-specific anchors.
Both radius-based and curvature-normalized representations exhibit high in-chain
coherence in these sequences.

Quantitatively, radius-based coordinates often achieve slightly lower local RMSE within
individual isotones, reflecting the fact that proton addition directly affects the
charge distribution measured by $R_{\mathrm{exp}}$.
Curvature-normalized coordinates, however, consistently yield higher $R^2$ and more
stable global trend shapes across elements.

These results indicate that mass scaling is complementary rather than dominant in
proton-driven geometry.
While curvature normalization does not substantially improve local fits in isotone
sequences, it does not degrade them and continues to provide a consistent geometric
reference across varying nuclear configurations.
\subsection{Mirror--pair test: negative control}

As a negative control, mirror nuclei $(Z,N)\leftrightarrow(N,Z)$ are examined where
experimental charge radii are available for both members of the pair.
Because charge radii are proton observables, mirror symmetry is not expected in
absolute terms, and Coulomb effects dominate the observed asymmetry.

When mirror asymmetries are expressed in both radius-based and curvature-normalized
coordinates, the relative differences in curvature space closely track those in radius
space.
No systematic amplification or suppression of proton--neutron asymmetry is observed
under mass scaling.

This result demonstrates that curvature normalization acts as a well-behaved coordinate
transformation that preserves known composition asymmetries rather than introducing
artificial structure.
The organizational advantages observed in isotopic chains and global analyses therefore
arise specifically in regimes where composition changes drive surface growth, and not
from a generic distortion of nuclear geometry.
Taken together, these validation tests establish that curvature normalization provides
a robust and selective improvement in nuclear geometry organization.
Its principal utility lies in isolating neutron-driven surface effects that are
suppressed in absolute-radius representations, while remaining consistent with
proton-driven geometry and known composition asymmetries.

	\section{Robustness Checks}
\label{app:robustness_checks}

A series of robustness checks was performed to ensure that the observed universal scaling is not an artifact of data selection, normalization choice, or statistical treatment.

\subsection{Light--Nucleus Exclusion}

As discussed in Sec.~\ref{sec:light_nucleus_validity}, nuclei with $Z \le 4$ occupy a few-body and clustering-dominated regime. When these nuclei are excluded from the analysis, the universal scaling remains unchanged while residual scatter tightens and extreme deviations collapse. This confirms that the scaling applies to bulk nuclear geometry and is not driven by light-nucleus behavior.

\subsection{Core--Isotope Selection}

The core isotope for each element is defined as the isotope with the smallest neutron number for which reliable charge-radius data exist. Alternative reasonable choices (e.g., neighboring stable isotopes) were tested and found to produce no qualitative change in the observed scaling or residual structure.

\subsection{Normalization Sensitivity}

The use of the mass-based normalization scale $\bar{\lambda}_C = \hbar/(mc)$ was tested against alternative normalizations based on physical radii alone. While alternative normalizations preserve qualitative trends within isotopic chains, they do not produce a comparable cross-element collapse. The observed universality is therefore specific to the dimensionless curvature normalization employed here.

\subsection{Binning and Visualization}

All figures were inspected under multiple binning and plotting choices. The three residual regimes described in Sec.~\ref{sec:residual_regimes} are robust under changes in bin width and plotting resolution. No features reported in the main text depend on a particular visualization choice.

	\section{Exploratory Overlay of Nuclear Decay Modes}
\label{app:decay_mode_overlay}

Although the present work focuses on neutron-skin geometry derived from experimental charge radii, it is instructive to examine how evaluated nuclear decay modes are distributed in the same normalized neutron-excess coordinate used throughout the analysis. This appendix presents an exploratory overlay of dominant decay modes drawn from the NuDat database~\cite{NuDat2023}, mapped into the coordinate
\[
x \equiv \frac{N_{\mathrm{excess}}}{Z},
\]
using the same element-specific core anchors $N_{\mathrm{core}}(Z)$ defined for the radii-based analysis.

\subsection{Method}

Ground-state nuclides and their dominant decay modes were extracted from the NuDat database maintained by the National Nuclear Data Center (NNDC)~\cite{NuDat2023}. NuDat compiles evaluated decay data derived primarily from the Evaluated Nuclear Structure Data File (ENSDF)~\cite{ENSDF}, which aggregates experimental results from $\beta$ decay, $\alpha$ decay, spontaneous fission, and related measurements.

For each nuclide with proton number $Z$ and neutron number $N$, the neutron excess was computed relative to the same $N_{\mathrm{core}}(Z)$ used in the neutron-skin analysis, and the corresponding value of $x$ was evaluated. No charge-radius data are required for this mapping.

Decay modes were grouped into broad, standard categories (e.g., $\beta^-$, EC/$\beta^+$, $\alpha$, proton emission, neutron emission, isomeric transition, spontaneous fission). When multiple decay channels are listed for a given nuclide, the dominant channel reported in NuDat was used. This overlay is intended solely as a descriptive diagnostic and does not enter into the neutron-skin scaling itself.

\subsection{Decay--Mode Fractions vs Normalized Neutron Excess}

\begin{figure}[t]
	\centering
	\includegraphics[width=0.85\linewidth]{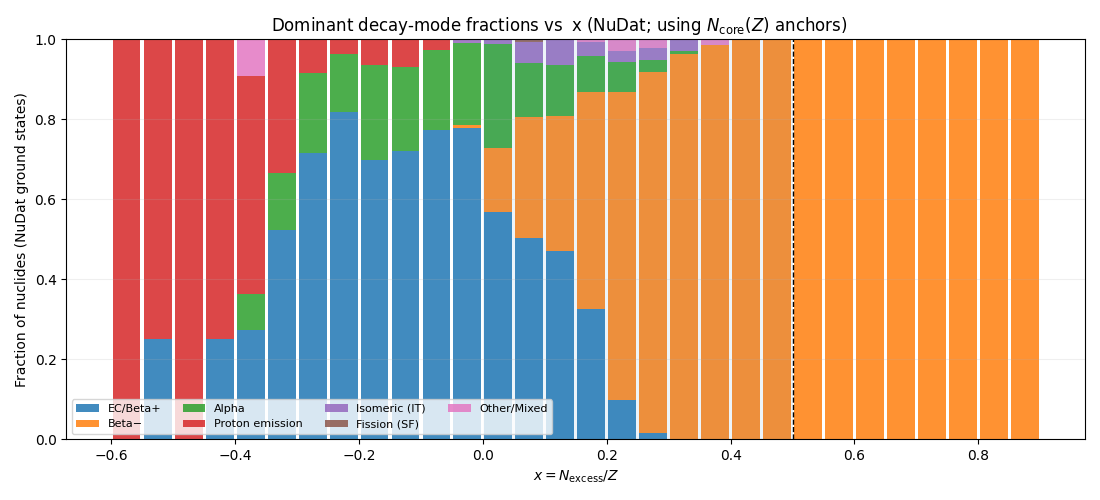}
	\caption{
		Dominant nuclear decay-mode fractions as a function of normalized neutron excess
		$x = N_{\mathrm{excess}}/Z$, computed for NuDat ground-state nuclides using the same
		core anchors $N_{\mathrm{core}}(Z)$ as the neutron-skin analysis.
		Bars show the fraction of nuclides in each $x$ bin exhibiting a given dominant decay
		channel.
		The shaded region ($x \le 0.5$) indicates the domain over which neutron-skin scaling
		is validated using experimental charge radii~\cite{AngeliMarinova2013}.
		Decay data are taken from the NuDat/ENSDF evaluated databases~\cite{NuDat2023,ENSDF}.
	}
	\label{fig:decay_mode_fractions}
\end{figure}

Figure~\ref{fig:decay_mode_fractions} shows the fraction of NuDat ground-state nuclides exhibiting each dominant decay mode as a function of $x$, binned uniformly. The shaded region indicates the domain $x \le 0.5$, which corresponds to the range over which the neutron-skin scaling is validated using experimental charge radii~\cite{AngeliMarinova2013}.

Several features are immediately apparent. Proton emission and EC/$\beta^+$ dominate at negative $x$, corresponding to proton-rich nuclides. As $x$ increases toward zero, $\alpha$ decay becomes prominent for heavy nuclei. For positive $x$ within the validated domain, $\beta^-$ decay is the dominant relaxation channel. Beyond $x \approx 0.5$, the distribution is almost entirely composed of $\beta^-$ and neutron-emission channels, reflecting the extreme neutron-rich regime documented in evaluated decay data~\cite{NuDat2023}.

\subsection{Extended Decay Map in \texorpdfstring{$(Z,x)$}{(Z, x)} Space}

\begin{figure}[t]
	\centering
	\includegraphics[width=0.85\linewidth]{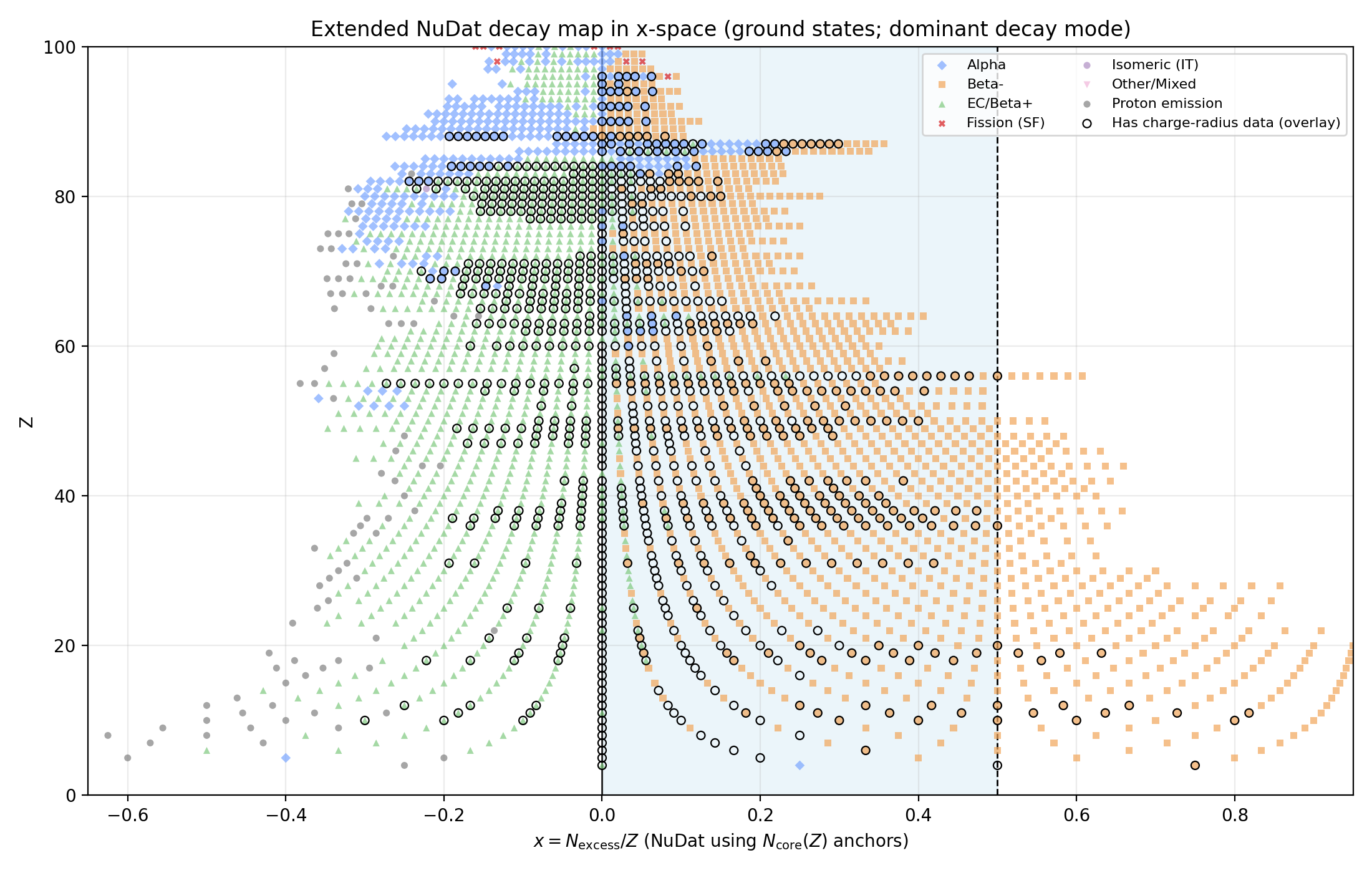}
	\caption{
		Extended NuDat decay map in the $(Z,x)$ plane, where
		$x = N_{\mathrm{excess}}/Z$ is computed using element-specific core anchors.
		Each point represents a NuDat ground-state nuclide, colored by dominant decay mode.
		Nuclides with experimentally measured charge radii are highlighted by overlaid markers.
		The dashed vertical line marks $x = 0.5$, separating the radii-validated domain from
		regions beyond it.
		Decay data are taken from the NuDat/ENSDF evaluated databases~\cite{NuDat2023,ENSDF}.
	}
	\label{fig:extended_decay_map}
\end{figure}
Figure~\ref{fig:extended_decay_map} presents the same decay information in the $(Z,x)$ plane. Each point corresponds to a NuDat ground-state nuclide, colored by dominant decay mode. Nuclides for which experimental charge-radius data exist are highlighted by an overlaid marker. The dashed vertical line marks $x=0.5$, separating the radii-validated domain from regions beyond it.

This representation shows that decay modes form structured bands in $(Z,x)$ space rather than being uniformly distributed. The set of nuclides with measured charge radii lies almost entirely within the validated domain and is dominated by weak decay and $\alpha$ decay channels. Spontaneous fission appears only at very high $Z$ and lies largely outside the domain of long-lived nuclides for which charge radii are available, consistent with known stability systematics~\cite{NuDat2023,ENSDF}.

\subsection{Interpretation and Scope}

These overlays demonstrate that dominant nuclear decay modes are non-uniformly distributed in the normalized neutron-excess coordinate used in the neutron-skin analysis. The ordering of decay modes across $x$ is consistent with established nuclear stability trends and illustrates that the same coordinate organizing neutron-skin geometry also provides a compact way to visualize decay-channel systematics.

We emphasize that this appendix is exploratory in nature. No claim is made that neutron-skin curvature or its residuals directly determine decay pathways, nor that decay modes can be predicted from the curvature framework alone. Rather, the purpose of this overlay is to document an empirical association between decay-channel prevalence and position in normalized nuclear-geometry space, and motivates further investigation of how such geometric organization constrains spectroscopic and response observables in atomic and molecular systems.

\section{Family--Conditioned Residual Structure and Geometry Branching}
\label{app:family_branching}

\subsection{Purpose and scope}
\label{app:family_branching_scope}
The main text establishes a global neutron-skin scaling backbone in the coordinate
\(
x \equiv N_{\mathrm{excess}}/Z
\)
with response
\(
y \equiv K_{R,\mathrm{skin}}/K_{R,\mathrm{core}}.
\)
Residual structure relative to this backbone is visibly non-random across the chart.
This appendix shows that these deviations are not ``noise,'' but are strongly organized
into a small number of geometry families and, in specific regions, split into multiple
stable branches. No chemical or shell labels are used as inputs to construct these
families; organization emerges from the curvature coordinate system itself.

This appendix has two goals:
(i) demonstrate that family conditioning sharply reduces residual scatter relative to the global backbone,
and (ii) identify where the chart exhibits branching behavior (multiple geometric paths) versus
where residual structure is dominated by parity modulation (odd--even effects).

\subsection{Definitions}
\label{app:family_branching_defs}

\paragraph{Backbone and residual.}
Let \(y_{\mathrm{backbone}}(x)\) denote the global backbone used in the main text (or its refined form),
and define the residual
\begin{equation}
	r_{\mathrm{global}} \equiv y - y_{\mathrm{backbone}}(x).
	\label{eq:r_global}
\end{equation}

\paragraph{Parity labels.}
We label neutron parity as
\(
\mathrm{odd\_even}_N \in \{\mathrm{even}, \mathrm{odd}\},
\)
with analogous proton and mass parity available when required.

\paragraph{Geometry families.}
We employ a novel nuclear-family nomenclature:
\(\mathrm{C1}\), \(\mathrm{C2}\), \(\mathrm{transitional}\), \(\alpha\)-trough, and \(\mathrm{C3\ onset}\),
plus a ``few-body'' out-of-domain label for very light nuclei.
For this appendix, the family label is treated operationally: nuclei are grouped into
families based on coherent behavior in \(r_{\mathrm{global}}\) and known transition zones
(\(\alpha\)-trough and C3 onset) established in prior analysis. For this paper this breakdown is not needed 
to demonstrate the presence of the global trend. 

\paragraph{Branching.}
Branching refers to the presence of two statistically distinct residual bands
within a restricted subgroup (e.g.\ a family and parity subset), indicating multiple
geometrically stable paths relative to the same backbone.
\pagebreak
\subsection{Three-panel summary figure}
\label{app:family_branching_fig}

\begin{figure}[H]
	\centering
	\includegraphics[width=\linewidth]{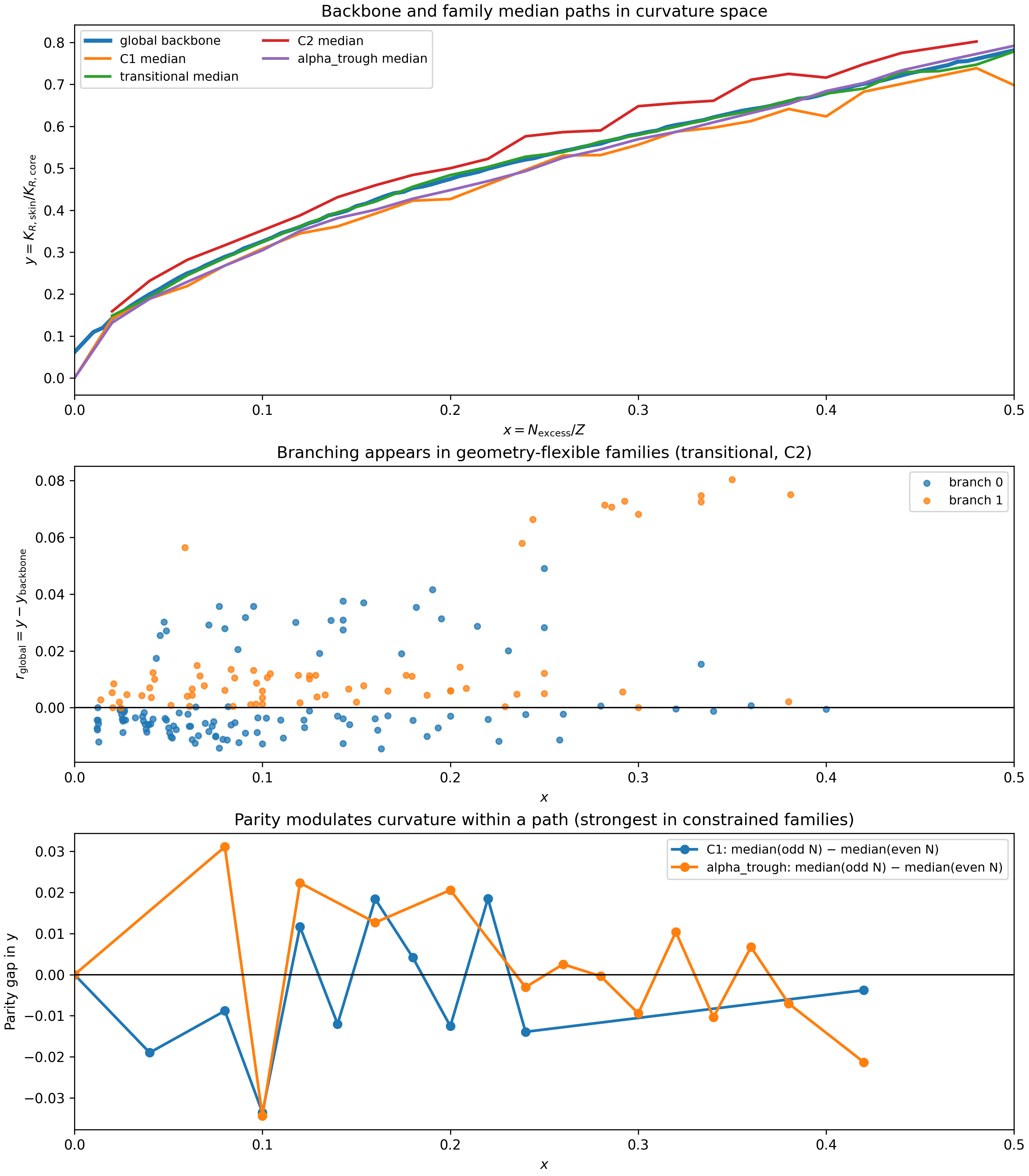}
	\caption{\textbf{Family-conditioned paths, branching, and parity modulation in curvature space.}
		\textbf{(a)} Global backbone \(y_{\mathrm{backbone}}(x)\) (thick line) with family-conditioned median
		paths in \((x,y)\). Families follow distinct geometric tracks around the backbone.
		\textbf{(b)} Residual branching relative to the backbone in transitional and C2 families, showing two
		separated residual bands even within fixed parity subsets.
		\textbf{(c)} Parity modulation within constrained families (C1 and \(\alpha\)-trough) shown as the binned
		median gap \(\mathrm{median}(y\,|\,\mathrm{odd}\ N)-\mathrm{median}(y\,|\,\mathrm{even}\ N)\).
	}
	\label{fig:family_branching_3panel}
\end{figure}

Figure~\ref{fig:family_branching_3panel} provides a compact visual summary of the result:

\begin{itemize}
	\item \textbf{Panel (a)} shows the global backbone along with family-conditioned median paths in the
	\((x,y)\) plane. Families occupy distinct tracks around the backbone, indicating different realizations
	of curvature growth with neutron excess.
	\item \textbf{Panel (b)} shows residual branching relative to the backbone in geometry-flexible families
	(transitional and C2). Two separated residual bands persist within fixed parity subsets, indicating
	branching is geometric and not solely pairing-driven.
	\item \textbf{Panel (c)} shows parity modulation (median odd--even neutron gap) in constrained families
	(C1 and \(\alpha\)-trough). Parity effects modulate curvature along a path but do not generate the
	path splitting seen in panel (b).
\end{itemize}

\subsection{Family--conditioned residual structure and branching statistics}
\label{app:family_branching_stats}

The global neutron-skin backbone introduced in the main text provides a universal
reference for curvature growth across the nuclear chart. However, residuals about
this backbone are visibly structured. To determine whether this structure reflects
random scatter or systematic nuclear geometry, we quantify how residual behavior
changes when nuclei are conditioned on curvature family, branch structure, and
parity.

Residual scatter is measured using a robust median absolute deviation (MAD) estimator,
computed for the global residual distribution and for each curvature family separately.
Values are reported relative to the global residual scatter, such that values
$\ll 1$ indicate strong collapse of residual structure once family geometry is
accounted for. Branching strength is quantified using the effect-size statistic
(Cohen's $d$) for bimodal residual distributions where branching is detected.
Parity modulation is summarized by the root-mean-square (RMS) odd--even neutron
gap in $y$ across $x$ bins.

\begin{table}[t]
	\centering
	\caption{\textbf{Family-conditioned residual collapse, branching, and parity modulation.}
		Residual scatter is reported as the ratio of family-conditioned MAD to the global
		residual MAD. Branching strength is summarized by a typical Cohen's $d$ value in
		subgroups where bimodal residual structure is detected. Parity modulation is
		summarized by the RMS of the binned median odd--even neutron gap.}
	\label{tab:family_branching_summary}
	\begin{adjustbox}{max width=\textwidth}
	\begin{tabular}{lrrrr}
		\toprule
		\textbf{Curvature family} &
		\textbf{$N$} &
		\boldmath{$\sigma_{\mathrm{MAD}}/\sigma_{\mathrm{MAD,global}}$} &
		\textbf{Branching (typ.\ $d$)} &
		\textbf{Parity gap RMS} \\
		\midrule
		C$_1$            & 423 & $\approx 0.00$ & none       & $0.0156$ \\
		Transitional     & 276 & $0.12$         & $\sim 3.0$ & $0.0106$ \\
		C$_2$            & 105 & $0.24$         & $\sim 5.5$ & $0.0269$ \\
		$\alpha$-trough  & 95  & $0.34$         & none       & $0.0165$ \\
		C$_3$ onset      & 28  & $0.50$         & emerging   & ---      \\
		\bottomrule
	\end{tabular}
	\end{adjustbox}
\end{table}

Table~\ref{tab:family_branching_summary} shows that conditioning on curvature family
accounts for the majority of residual structure observed in the global analysis.
In the baseline C$_1$ family, residual scatter collapses almost completely, indicating
that deviations from the backbone are dominated by odd--even (pairing) effects rather
than geometric diversity. In contrast, the transitional and C$_2$ families exhibit
strong residual collapse combined with statistically distinct branching, with large
effect sizes ($d \sim 3$--$5.5$) that persist within fixed parity subsets. This
demonstrates that branching reflects multiple geometrically stable curvature paths,
not pairing artifacts.

The $\alpha$-trough family exhibits suppressed curvature growth with moderate residual
collapse and enhanced parity sensitivity, consistent with a constrained geometric
regime. The C$_3$ onset region shows partial residual collapse with emerging structure,
reflecting the breakdown of simple curvature buffering and the onset of deformation
effects, though limited statistics preclude definitive branching classification.

Overall, the statistics in Table~\ref{tab:family_branching_summary} demonstrate that
residual structure in the main text is primarily geometric in origin, with branching
and parity acting as secondary modulators. This confirms that the curvature families
identified here are natural consequences of nuclear geometry rather than post hoc
classifications.

\subsection{Re--evaluating representative examples within their families}
\label{app:family_branching_examples}

The main text emphasizes global scaling across the chart. This appendix enables a sharper view by
re-evaluating representative examples \emph{within} their family context. The core pattern is:

\begin{itemize}
	\item In \textbf{C1} (baseline geometry), nuclei closely track the backbone and residuals are primarily
	parity-modulated. Apparent deviations in the global view often resolve into odd--even oscillations
	within a single path.
	\item In the \textbf{transitional} family, residuals separate into two stable branches over broad \(x\)
	ranges. Nuclei that appear as ``outliers'' in the global view often become inliers once assigned to
	the correct branch.
	\item In \textbf{C2}, the entire family occupies an elevated curvature-growth path relative to the backbone,
	with additional branch splitting in specific subregions. Pairing influences branch occupancy but does
	not generate the existence of branching.
	\item In the \(\boldsymbol{\alpha}\)\textbf{-trough}, curvature growth is suppressed and parity effects are enhanced;
	this regime is naturally understood as a constrained geometric channel rather than random scatter.
\end{itemize}

\subsection{Relation to Paper 2}
\label{app:family_branching_link_to_paper2}
The family-conditioned geometric paths identified here provide the structural foundation for the
spectroscopic manifestations analyzed in Paper 2, where nuclear geometry enters indirectly through
atomic and molecular observables. This appendix therefore serves as the bridge between global nuclear
scaling and downstream applications.

\subsection{Construction of nuclear family quadratic relations}
\label{app:quadratic_families}

Prior to introducing a global neutron-skin backbone, curvature growth was examined on an
element-by-element basis using only experimentally measured charge radii.
For each element with three or more isotopes, the outward curvature component
\(K_{R,\mathrm{skin}}\) was fitted empirically as a function of neutron excess,
\begin{equation}
	K_{R,\mathrm{skin}}(N_{\mathrm{excess}})
	= a_2 N_{\mathrm{excess}}^2 + a_1 N_{\mathrm{excess}} + a_0 ,
	\label{eq:quadratic_fit}
\end{equation}
with no parameters derived from theory and no assumptions regarding shell structure,
liquid-drop behavior, or force laws.

The quadratic form was selected solely on empirical adequacy. Across isotopic chains,
the data exhibit (i) smooth monotonic growth at small neutron excess, (ii) progressive
saturation as neutron buffering becomes less efficient, and (iii) eventual instability
or breakdown beyond the fitted region. Linear models could not capture saturation, while
higher-order polynomials were unnecessary and unstable over the available data range.
Equation~\eqref{eq:quadratic_fit} is therefore the lowest-order function capable of
describing the observed behavior.

\subsection{Geometric interpretation of the coefficients}

Each coefficient in Eq.~\eqref{eq:quadratic_fit} admits a consistent geometric
interpretation when viewed through the curvature-based framework employed in
this work:

\paragraph{Core reference offset (\texorpdfstring{$a_0$}{a\_0}).}

The constant term represents the curvature state at the anchor isotope. By construction,
\(a_0 \approx 0\) for most chains, reflecting that the nuclear core is a balanced
prior to the addition of neutron excess.

\paragraph{Linear growth rate (\texorpdfstring{$a_1$}{a\_1}).}

The linear coefficient measures the initial rate at which outward curvature is added per
neutron. Empirically, \(a_1\) characterizes the initial rate at which outward curvature
responds to neutron addition. Larger \(a_1\) corresponds to efficient early expansion, while smaller
\(a_1\) indicates stiff, strongly anchored curvature.

\paragraph{Quadratic stiffening (\texorpdfstring{$a_2$}{a\_2}).}

The quadratic coefficient is negative in stable regions and represents curvature saturation.
It captures the loss of marginal addition efficiency as neutrons accumulate.
The magnitude of \(a_2\) reflects saturation behavior in the curvature response,
aligning with shell filling and pairing effects contributing to the detailed residual
structure.

\subsection{Emergence of curvature families}

When the fitted coefficients \((a_1,a_2)\) are compared across elements, they do not form a
continuum. Instead, they cluster into a small number of regimes corresponding to distinct
curvature-growth behaviors. These regimes define the nuclear curvature families used
throughout this work:

\begin{itemize}
	\item \textbf{C$_1$ family:} modest \(a_1\), weak \(a_2\), yielding smooth, nearly universal
	quadratic behavior after normalization.
	\item \textbf{C$_2$ family:} larger \(a_1\) with stronger negative \(a_2\), producing elevated
	curvature growth that saturates more rapidly.
	\item \textbf{Transitional families:} intermediate and mixed coefficients, indicating
	reorganization of curvature topology.
	\item \textbf{$\alpha$-trough region:} suppressed quadratic growth with reduced \(a_1\) and
	small \(|a_2|\), consistent with deep inward anchoring.
	\item \textbf{C$_3$ onset:} irregular or noisy quadratic behavior, and less efficient neutron addition.
\end{itemize}

\subsection{Relation to normalized global scaling}

Upon normalization,
\(
y = K_{R,\mathrm{skin}}/K_{R,\mathrm{core}}
\)
and reparameterization by
\(
x = N_{\mathrm{excess}}/Z,
\)
these element-wise quadratics collapse into the global trend which can be further broken 
down into family-specific universal curves.
This demonstrates that the quadratic relations are not element-specific physics, but
express scale-invariant curvature trends. Differences between elements arise primarily
through family membership to a geometric structure and vertical offsets, rather than 
distinct functional forms.

Importantly, the same family structure reappears in the global-backbone residual and
branching analysis presented in Appendix~\ref{app:family_branching}, confirming that
the families are not artifacts of a particular fitting strategy but reflect reproducible 
curvature-response patterns in the nuclear data.

\subsection{Re-evaluation of quadratic results by curvature family}
\label{app:family_quadratic_reinterpretation}

The element-wise quadratic relations discussed here, and derivable by processes discussed earlier, were constructed
prior to the introduction of a global backbone or family framework.
With the benefit of the present family-conditioned analysis, those earlier results
can now be reinterpreted systematically.
Apparent variability or breakdown of quadratic behavior across elements corresponds
to curvature family membership and, in certain regions, to the presence of multiple 
geometrically consistent trajectories rather than statistical scatter.

\begin{table}[t]
	\centering
	\caption{\textbf{Empirical quadratic response and residual structure by curvature family.}
		Quadratic fits refer to atom-specific fits within isotopic chains. 
		Residual structure summarizes behavior relative to the global neutron-skin backbone.}
	\label{tab:family_empirical_structure}
	\begin{tabular}{>{\RaggedRight\arraybackslash}p{2.78cm}
			>{\RaggedRight\arraybackslash}p{6.2cm}
			>{\RaggedRight\arraybackslash}p{6.2cm}}
		\toprule
		\textbf{Family} &
		\textbf{Elemental Quadratics} &
		\textbf{Residual structure (this work)} \\
		\midrule
		
		\textbf{C$_1$ (baseline)} &
		Stable quadratic fits with modest linear growth and weak saturation terms. 
		Behavior becomes near-universal after normalization. &
		Residuals collapse tightly about the global backbone. 
		No branching detected. 
		Deviations are dominated by odd--even effects. \\
		
		\addlinespace
		
		\textbf{Transitional} &
		Element-dependent or distorted quadratic fits; increased coefficient variability; reduced goodness-of-fit. & Strong bimodal residual structure. Large effect sizes ($d \sim 3$). Branching persists within fixed parity subsets. \\
		
		\addlinespace
		
\textbf{Transitional} &
Element-dependent or distorted quadratic fits; increased coefficient variability; reduced goodness-of-fit. &
Strong bimodal residual structure; large effect sizes ($d \sim 3$); branching persists within fixed parity subsets. \\

\addlinespace

\textbf{C$_2$ (elevated)} &
Steeper early growth and stronger saturation; quadratic curves lie systematically above the C$_1$ baseline. &
Elevated residual path relative to the backbone; additional branching detected ($d > 5$ in subsets). \\

\addlinespace

\boldmath{$\alpha$}\textbf{-trough} &
Suppressed quadratic growth with reduced linear term and weak saturation; apparent flattening of response. &
No branching observed; strong and irregular parity sensitivity. \\

\addlinespace

\textbf{C$_3$ onset} &
Noisy or failing quadratic fits; breakdown of smooth saturation behavior. &
Partial residual collapse with emerging structure; limited statistics for branching analysis. \\
		\bottomrule
	\end{tabular}
\end{table}

\begin{table}[t]
	\centering
	\caption{\textbf{Geometric interpretation of curvature families.}
		Interpretations summarize the effective geometric regime implied by residual structure patterns.}
	\label{tab:family_geometric_interpretation}
	\begin{tabular}{p{3.5cm} p{12cm}}
		\toprule
		\textbf{Family} &
		\textbf{                                       Geometric interpretation} \\
		\midrule
		
		\textbf{     C$_1$ (baseline)} &
		Single, stable curvature-growth path. 
		Pairing acts as a secondary modulation rather than a structural driver. \\
		
		\addlinespace
		
		\textbf{        Transitional} &
		Geometry becomes flexible. 
		Multiple stable curvature realizations coexist across isotopic chains. \\
		
		\addlinespace
		
		\textbf{       C$_2$(elevated)} &
		Directional curvature growth with competing geometric channels. 
		Pairing influences branch occupancy but does not determine branch existence. \\
		
		\addlinespace
		
		\boldmath{              $\alpha$}\textbf{-trough} &
		Constrained geometry with deep inward anchoring. 
		Small neutron changes produce amplified geometric response. \\
		
		\addlinespace
		
		\textbf{        C$_3$ onset} &
		Breakdown of simple buffering regime. 
		Onset of deformation-driven geometric organization. \\
		
		\bottomrule
	\end{tabular}
\end{table}

\noindent\textbf{Summary.}
The family-conditioned interpretation shows that the quadratic found for individual atoms 
remain valid descriptors of curvature response \emph{within} their appropriate geometric 
regimes. Apparent deviations from simple quadratic behavior reflect either topology 
instability (branching) or constrained geometry, rather than failure of the underlying 
curvature-scaling framework.

	\FloatBarrier
	\section*{Acknowledgements}
	\addcontentsline{toc}{section}{Acknowledgements}
	
	The author thanks the scientific community for maintaining open access to the
	empirical datasets that enabled this work, and the contributors and reviewers
	who provided feedback on clarity and presentation.
	No external funding was received for this research.
	
	Analyses and figures were produced using open-source scientific software and are
	intended to be reproducible from the referenced datasets and scripts.
	AI-assisted tools were used in a limited capacity for editorial feedback,
	data collection, code review, and assistance with language clarity.
	All scientific analysis, interpretation, and conclusions are the sole
	responsibility of the author.

	\FloatBarrier
	\clearpage
	\bibliographystyle{unsrt}
	\bibliography{references}
	
\end{document}